\documentclass[epsfig,12pt]{article}
\usepackage{graphicx}
\usepackage{epsfig}
\usepackage{graphicx}
\usepackage{amssymb}
\usepackage{amsmath}
\usepackage{mathtools}
\usepackage{hyperref}
\usepackage{subcaption}
\usepackage{comment}
\usepackage{float}
\usepackage{tikz}
\usetikzlibrary{shapes}
\usetikzlibrary{positioning}
\usetikzlibrary{chains}
\usetikzlibrary{arrows,fit,decorations.pathreplacing, shapes.misc}
\usetikzlibrary{decorations.markings}
\tikzstyle{every picture}+=[remember picture]
\tikzstyle{na} = [baseline=-.5ex]
\usepackage{tikz-feynman}

\usepackage[nottoc,notlof,notlot]{tocbibind} 
\usepackage[titles]{tocloft}

\usepackage{array}

\usepackage{todonotes}

\newcommand{\beq}{\begin{equation}}   
\newcommand{\eeq}{\end{equation}}
\newcommand{\beqn}{\begin{eqnarray}}   
\newcommand{\eeqn}{\end{eqnarray}}

\newcommand{\mtr}[3]{#1_{#2\bar{#3}}\,}

\newcommand{\riec}[4]{R_{\bar{#1}#2#3}^{~ ~ ~ #4}}

\begin{document}
\unitlength = 1mm

\def\de{\partial}
\def\Tr{ \hbox{\rm Tr}}
\def\const{\hbox {\rm const.}}  
\def\o{\over}
\def\im{\hbox{\rm Im}}
\def\re{\hbox{\rm Re}}
\def\bra{\langle}\def\ket{\rangle}
\def\Arg{\hbox {\rm Arg}}
\def\Re{\hbox {\rm Re}}
\def\Im{\hbox {\rm Im}}
\def\diag{\hbox{\rm diag}}
\def\mtr#1#2#3{#1_{#2\bar{#3}}}

\def\mc{\mathcal}
\def\dx{{\rm d}^2x}
\def\dt{{\rm d}\theta^+}
\def\dtb{{\rm d}\bar{\theta}^+}
\def\mf{\mathfrak}
\def\X{\!\!&\times&\!\!}
\def\={\!\!&=&\!}
\def\+{\!\!&+&\!}
\def\-{\!\!&-&\!}
\def\eqq{\!\!&\equiv&\!}
\def\th{\theta^+}
\def\thb{\bar{\theta}^+}


\def\Tr{{\rm Tr}}
\def\res{{\rm res}}
\def\Bf#1{\mbox{\boldmath $#1$}}
\def\balpha{{\Bf\alpha}}
\def\bbeta{{\Bf\beta}}
\def\bgamma{{\Bf\gamma}}
\def\bnu{{\Bf\nu}}
\def\bmu{{\Bf\mu}}
\def\bphi{{\Bf\phi}}
\def\bPhi{{\Bf\Phi}}
\def\bomega{{\Bf\omega}}
\def\blambda{{\Bf\lambda}}
\def\brho{{\Bf\rho}}
\def\bsigma{{\bfit\sigma}}
\def\bxi{{\Bf\xi}}
\def\bbeta{{\Bf\eta}}
\def\d{\partial}
\def\der#1#2{\frac{\d{#1}}{\d{#2}}}
\def\Im{{\rm Im}}
\def\Re{{\rm Re}}
\def\rank{{\rm rank}}
\def\diag{{\rm diag}}
\def\2{{1\over 2}}
\def\ntwo{${\mathcal N}=2\;$}
\def\nfour{${\mathcal N}=4\;$}
\def\none{${\mathcal N}=1\;$}
\def\ntwot{${\mathcal N}=(2,2)\;$}
\def\ntwoo{${\mathcal N}=(0,2)\;$}
\def\x{\stackrel{\otimes}{,}}
\def\wcpt{\mathbb{WCP}(2,2)}
\def\pd{\partial}
\newcommand{\ttr}[3]{#1_{#2\overline{#3}}}
\def\tt{\tau}
\def\vp{\varphi}
\def\hh{\sqrt{a^{2}+4r^{2}}}
\def\tg{\tilde{g}}
\def\mtr#1#2#3{#1_{#2\bar{#3}}}
\def\ml#1{\marginpar{\textcolor{red}{\footnotesize #1}}}

\def\ntwo{${\mathcal N}=2\;$}
\def\nfour{${\mathcal N}=4\;$}
\def\none{${\mathcal N}=1\;$}
\def\ntwot{${\mathcal N}=(2,2)\;$}
\def\ntwoo{${\mathcal N}=(0,2)\;$}

\newcommand{\cpn}{CP$(N-1)\;$}
\newcommand{\wcpn}{wCP${N,\widetilde{N}}(N_f-1)\;$}
\newcommand{\wcpd}{wCP$_{\widetilde{N},N}(N_f-1)\;$}
\newcommand{\WCP}{$\mathbb{WCP}(N,\tilde N)\;$}

\newcommand{\pt}{\partial}
\newcommand{\tN}{\widetilde{N}}
\newcommand{\ve}{\varepsilon}
\newcommand{\vr}{\varrho}
\renewcommand{\theequation}{\thesection.\arabic{equation}}

\newcommand{\sun}{SU$(N)\;$}

\setcounter{footnote}0

\vfill

\begin{titlepage}

\vspace{-1cm}

\begin{flushright}
FTPI-MINN-20-16, UMN-TH-3918/20\\
\end{flushright}

\vspace{2mm}

\begin{center}
{  \Large \bf  
Long Way to Ricci Flatness 
}

\vspace{4mm}

{\large \bf   Jin Chen$^a$, Chao-Hsiang Sheu$^{b}$, Mikhail Shifman$^{b,c}$, Gianni Tallarita$^{d}$ and 
Alexei Yung$^{e,c}$}
\end {center}

\begin{center}

{\it $^{a}$ Yau Mathematical Sciences Center, 
Tsinghua University,
Haidian, Beijing, 100084, China
}\\
    {\it  $^{b}$Department of Physics,
University of Minnesota,
Minneapolis, MN 55455}\\
{\it  $^{c}$William I. Fine Theoretical Physics Institute,
University of Minnesota,
Minneapolis, MN 55455}\\
{\it $^{d}$Departamento de Ciencias, Facultad de Artes Liberales, Universidad Adolfo Ib\'a\~nez,
Santiago 7941169, Chile
}\\
$^{e}${\it National Research Center ``Kurchatov Institute'', 
Petersburg Nuclear Physics Institute, Gatchina, St. Petersburg
188300, Russia}\\
\end{center}

\vspace{8mm}

\begin{center}
{\large\bf Abstract}
\end{center}

We study  two-dimensional weighted \ntwot supersymmetric $\mathbb{CP}$ models with the goal of exploring their infrared (IR) limit.   $\mathbb{WCP}(N,\tN)$ are simplified versions of world-sheet theories on  non-Abelian strings in four-dimensional \ntwo QCD. In the gauged linear sigma model (GLSM)  formulation,
$\mathbb{WCP}(N,\tN)$   has $N$ charges +1 and $\tN$ charges $-1$ fields. As well-known, at $\tN=N$ this GLSM is conformal. Its target space is believed to be a non-compact Calabi-Yau manifold. We mostly focus on the $N=2$ case, then the  Calabi-Yau space is a conifold. 

On the other hand, in 
the non-linear sigma model (NLSM) formulation the model has ultra-violet logarithms and does not look conformal.
Moreover, its metric is not Ricci-flat. We address this puzzle by studying the renormalization group (RG) flow 
of the model. We show that the metric of NLSM becomes Ricci-flat in the IR. Moreover, it tends to the known
 metric of the resolved conifold. We also study a close relative of the  $\mathbb{WCP}$ model -- the so called $zn$ model -- which in actuality represents the world sheet theory on a non-Abelian semilocal string and show that this $zn$ model has 
similar RG properties.

%

\end{titlepage}

\newpage

\section{Introduction}
\label{intro}

Two-dimensional $\mathbb{CP}(N-1)$ models gained a renewed attention recently because they arise as  world sheet theories on  non-Abelian strings in four-dimensional gauge theories. Non-Abelian vortex strings were first found in \ntwo  supersymmetric QCD (SQCD)
with the gauge group U$(N)$ and $N_f = N$ flavors of quark hypermultiplets \cite{HT1,ABEKY,SYmon,HT2}. In addition to four translational moduli, the  non-Abelian vortices  have  orientational moduli.  Their low-energy dynamics is described by two-dimensional  \ntwot supersymmetric    $\mathbb{CP}(N-1)$ model   on the string world sheet, see  \cite{Trev,Jrev,SYrev,Trev2} for reviews. 

If the number of quark flavors in four-dimensional \ntwo QCD exceeds the number of colors, $N_f > N$ the world sheet theory becomes what is usually referred to in the physical literature as a weighted $\mathbb{CP}$ ($\mathbb{WCP}(N,\tilde N)$)  model\,\footnote{In fact, $\mathbb{WCP}(N,\tilde N)$ is a simplified version 
of the world sheet theory on  semilocal strings. The actual world sheet theory is given by so called $zn$ model
\cite{SVY}. We will discuss both types of models in this paper.}  
\cite{HT1,HT2,SYsem,Jsem,SVY}, where $\tN=N_f-N$. 

A transparent formulation of $\mathbb{WCP}(N,\tilde N)$  was suggested by Witten \cite{W79,W2} (see also \cite{HaHo,V}) in terms of a gauged linear sigma model (GLSM). In this formulation $\mathbb{WCP}(N,\tilde N)$ is considered as a low-energy limit
on the Higgs branch of a U(1) gauge theory (supersymmetric QED with the Fayet-Iliopouls term)  with matter superfields: $N$ of them with charge $+1$ are denoted by $n_i$ and 
$\tilde N$ with charge $-1$ are denoted by $\rho_a$. The $\mathbb{WCP}(N,\tilde N)$ target space  can be obtained by integrating out the gauge multiplet which acquires a large mass $M_V$ due to the Higgs mechanism.

We will focus on a special case $N=\tilde N$ which is of a particular importance for the dynamics of the non-Abelian strings. In 
this case  the world sheet $\mathbb{WCP}(N,\tN)$ model in the GLSM formulation becomes conformal. The only ultraviolet-divergent logarithm  appears in the renormalization of the Fayet-Iliopoulos (FI) parameter $\beta$ of the model.  It is exhausted by a single tadpole graph proportional to the difference in the numbers of positive and negative charges, i.e. $(N-\tN)$,  and vanishes at $N=\tilde N$. It is believed that the target space of $\mathbb{WCP}(N, N)$ model  reduces to a non-compact
Calabi-Yau manifold, equipped with a Ricci-flat metric (see \cite{NV}). The latter implies that the beta function in the model must vanish,
\beq
\beta_{ij}(g) \sim R_{ij} = 0\,.
\label{11}
\eeq

A particularly interesting case is $N=\tN =2$. As was shown in \cite{SYcstring,KSYconifold,SYlittles},  if $N=\tN =2$  the non-Abelian vortex behaves as a critical superstring.  This happens because in this case four translational moduli of the non-Abelian vortex
combined with orientational and size moduli form a ten-dimensional space required for a superstring to become critical.
The target space of our $\mathbb{WCP}(N, N)$ model in this case  becomes six-dimensional Calabi-Yau space, the conifold,
see \cite{NV} for a review. In this paper we mostly focus on the conifold case.

The above considerations come in  contradiction with the analysis in the NLSM formulation
of the $\mathbb{WCP}(N,N)$ model. In the latter approach one assumes the Higgs regime in the U(1) gauge theory and 
 uses classical equations of motion to eliminate heavy gauge and Higgs fields at energies  $\ll M_V$ neglecting their kinetic terms.  Then it turns out that   the model has ultra-violet logarithms of the type $\log M_V/\mu$ where $\mu$ is an IR scale.
Moreover, its metric is not Ricci-flat so its beta function does not vanish \cite{SVY,SS}. The model is not apparently conformal.

It is important that in the case of $\mathbb{CP}(N-1)$ models associated with a compact target space this contradiction does not occur;
$\mathbb{CP}(N-1)$ model in both GLSM and NLSM formulations has the same beta function.

A similar  puzzle was noted recently \cite{ARSW} in the simplest case of $\mathbb{WCP}(1,1)$.  In this case duality arguments suggest that the model should be a free field theory in the IR while the NLSM formulation gives  a {\em non-trivial} Ricci tensor. A numerical solution of the renormalization group (RG) equations in  \cite{ARSW}  shows that the solution in fact flows to a free theory in the IR.

In this paper we generalize this idea to the  $\mathbb{WCP}(2, 2)$ model. The desired IR limit now is not a free theory, but, rather Ricci-flat.  We study the RG flow  in the  $\mathbb{WCP}(2, 2)$ case and demonstrate that the NLSM metric indeed approaches the Ricci-flat  conifold solution of   \cite{Candelas:1989js,PandoZayas:2000ctr} in the IR.

Next we analyze the $zn$ model which actually represents the world sheet theory on the non-Abelian string \cite{SVY} and 
show that it has a similar RG flow.

Our qualitative understanding of this result is as follows. As a warm-up let us start with the $\mathbb{CP}(N-1)$ model. The NLSM formulation assumes the Higgs regime.  One component of a charge multiplet of fields $n_i$, $i=1,...,N$, say, $n_2$ in Secs. \ref{asoln}, \ref{geoboson} or $n_1$ in Sec. \ref{sec4},  develops a vacuum expectation value (VEV). It becomies massive while $(N-1)$ other components are massless Goldstone fields fluctuating over the target space. The global SU$(N)$ symmetry of the model is not realized linearly in the NLSM Lagrangian. 

This classical picture does not survive at the quantum level as was shown by Witten long ago \cite{W79,W2}. In the solution obtained at the quantum level the fields $n_i$  develop no VEVs, they are smeared all over the
target space  of the model. All fields $n_i$ acquire mass gap and the SU$(N)$ global symmetry is restored.
At the very end both formulations, GLSM and NLSM, arrive at one and the same solution.\footnote{Say, the CP(1) model in the NLSM formulation was solved in \cite{zam} long ago; this solution exhibits the same features as Witten's GLSM description. }

The lesson to learn from this is that the NLSM setup is not ``transparent" in a sense that it starts from a picture very distant from the final solution.
It  ignores ``microscopic" physics captured by  GLSM. The final IR results are the same, 
but the NLSM road to it is not so straightforward as the GLSM one. This is especially true for supersymmentric models in which GLSM 
allows one to apply such powerful methods as large $N$ expansion \cite{W79} and exact twisted superpotentials \cite{W2}.

In the $\mathbb{WCP}(N,N)$ model we also expect that the charged fields after all have no VEVs.
At  the quantum level $n$ and $\rho$ fields are smeared all over the non-compact Higgs branch.\footnote{The model has also the Coulomb branch. It opens up at the value of the FI parameter
$\beta = 0$. We consider nonvanishing $\beta$ in this paper.} 
The NLSM formulation gives us a bad starting point. The road from this starting point to the IR answer is non-trivial.
And still, one can reach the desired endpoint, as will be shown below.

The paper is organized as follows. In Sec.~\ref{model} we present $\mathbb{WCP}(N,N)$ model and discuss general aspects of the RG procedure. In Sec.~\ref{asoln} we review the Calabi-Yau metric on the conifold.  In Sec.~\ref{geoboson} we study the 
RG flow of the $\mathbb{WCP}(N,N)$ model in the NLSM formulation, while  in Sec.~\ref{numer} we present our numerical solution of the RG equations. In Sec.~\ref{sec51} we study the vacuum structure of $\mathbb{WCP}(N,N)$ model using  the exact twisted superpotential.    Section~\ref{sec4} is devoted to the emerging $Z$ factors in NLSM. In Sec.~\ref{zn} we consider the RG properties of the $zn$ model.

\section{The \boldmath{$\mathbb{WCP}(N,N)$} model}
\label{model}
\setcounter{equation}{0}

Let us  present the \ntwot supersymmetric $\mathbb{WCP}(N,N)$ model using the GLSM formulation.
First, we introduce two types (or flavors) of complex fields $n_k$ and $\rho_a$, 
with the electric charges $+1$ and $-1$, respectively,
\beqn
S
& =&
\int d^2 x \left\{
 \left|\nabla_{\mu} n_{k}\right|^2 +  \left|\tilde\nabla_{\mu} \rho_{a}\right|^2  + \frac1{4e^2}F^2_{\mu\nu} + \frac1{e^2}
|\pt_\mu\sigma|^2+\frac1{2e^2}D^2
\right.
\nonumber\\[3mm]
 &+&    2|\sigma|^2\left(  |n_{k}|^2  + |\rho_a|^2\right) + iD \left(|n_{k}|^2 - |\rho_a|^2 -\beta\right)
\Big\} +\mbox{fermions}\,.
\label{cpg}
\eeqn
Both indices $k$ and $a$ are integers  running from $1$ to  $N$ in the case under consideration. The action above is written in Euclidean conventions. 
The parameter $\beta$ in the last term of Eq. (\ref{cpg}) is dimensionless. It represents the two-dimensional Fayet-Iliopoulos term. 

The U(1) gauge field $A_\mu$ acts 
on $n$ and $\rho$ through appropriately defined covariant derivatives,
\beq\label{22}
 \nabla_{\mu}=\d_{\mu}-iA_{\mu}\,, \qquad \widetilde{\nabla}_{\mu}=\d_{\mu}+iA_{\mu}\,,
 \eeq
reflecting the sign difference between the charges. The fields $A_{\mu}$, complex scalar $\sigma$ and auxiliary real field $D$
form the bosonic part of the U(1) gauge supermultiplet. The electric coupling constant $e^2$ has dimension of mass squared. It is supposed to be large. The last term ($D$-term) classically enforces condensation of charged fields. In the Higgs phase
the gauge multiplet becomes massive. The scale of the gauge fields mass  is defined through the
product
\beq
M^2_V = 2e^2 \beta\,.
\label{23t}
\eeq

At energies much below $M_V$ all heavy fields (i.e. 
the gauge and Higgs supermultiplets) 
can be integrated out, and we are left with the low-energy sigma model on the Higgs branch. All terms except the kinetic terms of $n$ and $\rho$ disappear from the action, while the last term
reduces to the constraint
\beq
\sum_{k=1}^N |n_{k}|^2 - \sum_{a=1}^{{N}} |\rho_a|^2 =\beta\,.
\label{15}
\eeq
This constraint defines the   (real) dimension of the Higgs branch
\beq
{\rm dim}\,{\cal H} = 2(N+\tN -1) = 2(2N-1),
\label{dimH}
\eeq
where $2N$ and $2\tN$ are numbers of real degrees of freedom of fields $n_k$ and $\rho_a$ respectively, while $-2$ is  associated with the real constraint \eqref{15} and   one phase eaten by the  Higgs mechanism. For the conifold case 
$N=2$ we have  ${\rm dim}\,{\cal H}=6$.

The global symmetry of the model \eqref{cpg} is
\beq
 {\rm SU}(N)\times {\rm SU}(N)\times {\rm U}(1).
\label{globalgroup}
\eeq

\begin{figure}[!t]
\centerline{\includegraphics[width=4cm]{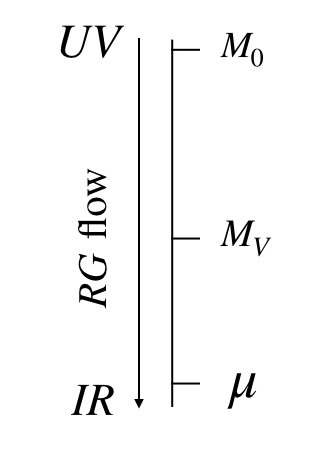}}
\caption{\small Three scales relevant for the RG flow of (\ref{cpg}.}
\label{pic1} 
\end{figure}

The RG flow domain we are interested in is depicted in Fig. \ref{pic1}. Here $M_0$ is the genuine UV scale where the action (\ref{cpg})
is formulated, $M_V$ is the parameter defined in (\ref{23t}) while $\mu $ is the sliding renormalization point.

We start our consideration of the  RG flow at $\mu =M_0$. Until we reach $\mu=M_V$ there is no flow. The only parameter which could be renormalized is 
the Fayet-Iliopoulos parameter $\beta$. However, the contributions of the fields $n$ and $\rho$ cancel each other due to the fact that the signs in the last term in  (\ref{cpg}) are opposite (see e.g.\cite{W2}
\footnote{In our previous works notation was different. The Fayet-Iliopoulos (FI) term $\beta$ in (\ref{15}) was denoted as $r$ in \cite{SS}. In the latter paper the coordinate patch was chosen to be the one with $\rho_{N}\neq 0$. In the present paper, we adopt, instead,  a dual patch with non-vanishing $n_{N}$, namely, $n_N =\sqrt\beta$ which interchanges the role of $z$ and $w$ compared to \cite{SS} and flips the sign of the FI term. In the present paper $\beta >0$. Note, however, that in Sec. \ref{sec4} 
we use the patch $n_1 = \sqrt\beta$ for technical reasons. }). 

Situation changes once we cross the line $M_V$ on Fig. \ref{pic1}. Once $\mu \ll M_V$ we cross into the domain of NLSM, with the gauge multiplet fields $A_{\mu}$, $D$ and $\sigma$ integrated out (their mass is represented by $M_V$).
 The target spaces in these cases are non-Einsteinian noncompact manifolds. Hence, these models are not renormalizable in the conventional sense of this word. 
 
 Discussion of some previous results in the \WCP model  which inspired the current work can be found in \cite{KS,ARSW}.  In the latter paper the simplest case $N=\tilde N =1$ was analyzed. Here we will address the general situation with arbitrary $N$ focusing mostly on the case $N=\tilde N =2$.
 
Transition from GLSM in Eq. (\ref{cpg}) to NLSM below $M_V$ was analyzed in detail for arbitrary $N$ and $\tilde N$ in Ref. \cite{SS}. Renormalization of the effective action proves to be rather complicated. At one loop it appears in the form of corrections containing logarithms
\beq
\log \frac{M_V}{\mu}
\label{116}
\eeq
due to $Z$ factors of the fields $n$ and $\rho$. Since the $Z$ factors are not protected and their RG flow is not limited to one loop, each subsequent loop
adds its own correction, see e.g. \cite{SS}. Note that the logarithm in (\ref{116}) differs from the standard UV/IR logarithm $\log M_0/\mu$. They coincide only in the limit
$M_V\to M_0$. This is in one-to-one correspondence with the fact that renormalization comes from the $Z$ factors. 

Our strategy is to write the RG equations 
in an appropriate {\em Ansatz}, determine the boundary condition in the ``UV'' and then analyze the RG flow in the IR to demonstrate that in this limit the Ricci tensor tends to zero. The ``UV" above is in the quotation marks because it refers to the scale $M_V$ which does not necessarily coincide with $M_0$. At  the scale $M_V$ the Ricci tensor 
is not flat at all. In this way we generalize the simplest case $N=\tilde N=1$ analyzed in \cite{ARSW} where the Ricci tensor is one-component and the IR flow indeed 
tends to make it approximately zero. The latter paper is titled ``A Long Flow to Freedom" which explains the choice of our title.

\section{The metric of the resolved conifold}
\label{asoln}
\setcounter{equation}{0}

In this section we review the metric on the conifold found in \cite{Candelas:1989js,PandoZayas:2000ctr}. Conifold can be defined as a Higgs branch of the GLSM \eqref{cpg} for $N=2$ subject to the constraint \eqref{15}.
Let us   construct  the U(1) gauge-invariant ``mesonic'' variables from the fields $n$ and $\rho$,
\beq
M_{ia}= n_i \rho_a.
\label{M}
\eeq
These variables are subject to the constraint
\beq
{\rm det}\, M_{ia} =0.
\label{coni}
\eeq
The matrix $M_{ia}$ has four complex parameters so the above  equation defines threefold in $\mathbb{C}^4$ in accordance with
\eqref{dimH}.

Equation (\ref{coni}) together with the requirements that the metric of the manifold should be K\"ahler (this is ensured by 
\ntwot supersymmetry of GLSM \eqref{cpg}) and Ricci-flatness
defines the non-compact Calabi-Yau space known as  conifold \cite{Candelas:1989js},  see also \cite{NV} for a review.
It is a cone which can be parametrized 
by the non-compact radial coordinate 
\beq
r^2={\rm Tr}\, M M^\dagger\,
\label{r}
\eeq
and five angles, see \cite{Candelas:1989js}. Its section at fixed $r$ is $S_2\times S_3$.

At $\beta =0$ the conifold develops a conical singularity, so both $S_2$ and $S_3$  
can shrink to zero. The explicit metric of the singular conifold was found in \cite{Candelas:1989js}.
Large values of $\beta$ correspond to weak coupling.

One way to smoothen the conifold singularity is by deforming its K\"ahler form.  This option is called the resolved conifold and amounts to introducing  a non-zero $\beta$ in (\ref{15}). This resolution preserves 
the K\"ahler structure and Ricci-flatness of the metric. 
If we put $\rho_a=0$ in (\ref{15}) we get the $\mathbb{CP}(1)$ model with the sphere $S_2$ of the  radius $\sqrt{\beta}$ as 
a target space. Thus, $S_2$ cannot shrink to zero at positive  $\beta$.

 The explicit form of the metric on the resolved conifold was found in \cite{PandoZayas:2000ctr}. Noting that for K\"ahler
manifolds the metric is given by 
\beq
g_{i\bar{j}} = \pt_i \pt_{\bar{j}} K,
\eeq
where $K$ is the K\"ahler potential the authors of \cite{PandoZayas:2000ctr} look for the solution of the  Ricci-flatness condition
using the {\em Ansatz}
\beq
	K = f(r^{2}) + \beta \log\left( 1 + \frac{ |n_1|^{2}}{ |n_2|^{2}}\right),
\label{ansatz}
\eeq
in the patch where $n_2\neq 0$. Here $f(r^2)$ is a function of the radial coordinate \eqref{r}. The motivation for this {\em Ansatz} is as follows. First note, that both terms here are invariant with respect to the global symmetry group \eqref{globalgroup} \footnote{ To check this invariance for the second term above note, that the  K\"ahler potential is defined up to an additional  holomorphic or anti-holomorphic function.}. Moreover, it turns out that the metric associated with the first term in \eqref{ansatz} vanishes at $r=0$ on the Ricci-flat solution, while the second term produces the round Fubini-Study CP(1) metric for the sphere $S_2$ of the radius $\sqrt{\beta}$ as expected for the resolved conifold \cite{Candelas:1989js,PandoZayas:2000ctr}. The very same  {\em Ansatz} naturally emerged in the perturbative analysis in \cite{SS}.

For the K\"ahler manifolds the Ricci tensor is given by the formula
\beq
{R}_{i\bar j} = -\pd_{i}\pd_{\bar{j}} \log{\det \mtr{g}{k}{l}}\,.
\eeq
Using the {\em Ansatz} \eqref{ansatz} one gets 
\beq
	\det(\mtr{g}{i}{j})  
	=  f' \, \left( \beta + r^{2} f' \right) \left( f' + r^{2} f'' \right)
\eeq
where prime denotes   {\em derivative with respect to} $r^{2}$. Then the condition of Ricci-flatness
leads to the equation
\begin{equation}\label{21}
	 f' \left( \beta + r^{2} f' \right) \left( f' + r^{2} f'' \right) = \frac{2}{3}
\end{equation}
or
\beqn
&&\gamma' \gamma (\beta +\gamma) 
= \frac{2}{3} r^2\,, \nonumber\\[3mm] 
&&\gamma(r^2) \equiv r^2 f'(r^2).
\label{23}
\eeqn
In Eqs. \eqref{21}, \eqref{23} the integration constant $2/3$ is chosen to fix the overall scale of $M_{ia}$ in \eqref{coni} and hence the  scale of the radial coordinate $r$, see below. 

Imposing the boundary condition 
\begin{equation}\label{33}
	\gamma \left(r^{2}\right)\Big|_{r^{2}=0} = 0
\end{equation}
to match the limit $\beta \to 0$ of the singular  conifold \cite{Candelas:1989js}, we can integrate (\ref{23}) to get
\footnote{Note that  for $N>2$, the algebraic equation (\ref{algebraic}) to determine $\gamma$ is of degree five or higher and  has no analytic solution.}
\begin{equation}\label{algebraic}
		\gamma^3 + \frac32\beta\gamma^2 -r^4 =0\,.
\end{equation}
It is not difficult to solve this equation -- we pick up the only real solution, namely,
\begin{equation}\label{24}
	\gamma_{\ast}(r^{2})  =
	-\frac{1}{2}\beta + \frac{1}{4}\beta^{2}\nu^{-1/3}(r^{2}) + \nu^{1/3}(r^{2})
\end{equation}
where
\begin{equation}
	\nu(r^{2}) = \frac{1}{2}\left[ r^{4} - \frac{1}{4}\beta^{3} + \left( r^{8}-\frac{1}{2}\beta^{3}r^{4} \right)^{1/2} \,\right] \,.
\end{equation}
and the subscript ${\ast}$ denotes the Ricci-flat solution.
This solution matches the  boundary condition (\ref{33}) provided we pick up the phase for $\nu(r^{2})^{1/3}$ 
equal  to $e^{i\pi/3}$ at the origin $r^2=0$. Also note, that with the  scale of $r$ fixed as in \eqref{21} (by the choice of the integration constant equal $2/3$) the solution for  $\gamma_{\ast}$ behaves as  $\gamma_{\ast} = r^{4/3}$ with the unit coefficient in the limit  $\beta\to 0$, see \cite{Candelas:1989js}.

 We conclude this section noting that $f_{\ast}(r^{2})$ can be written down explicitly as
\begin{equation}\label{27}
	f_{\ast}(r^{2}) = \frac{3}{2} \left[ \gamma_{\ast} - \frac{\beta}{2}\log\left( 3+\frac{2\gamma_{\ast}}{\beta} \right) \right]
\end{equation}
where $\gamma_{\ast}$ is defined in (\ref{24}). 




\section{Renormalization group flow of \boldmath{$\mathbb{WCP}\boldmath{(N,N)}$} }
\label{geoboson}
\setcounter{equation}{0}

To obtain the renormalization group equation in the $\mathbb{WCP}\boldmath{(N,N)}$ model, first recall the NLSM  formulation of the  model at hand,  \eqref{cpg}. To do so, we must take into account that (as we have already explained in Sec.~\ref{model}) the constraint (\ref{15}) and the U(1) gauge invariance
reduce the number of complex fields from $2N$ in the set $\{n_i\}+ \{\rho_a\}$  down to $2N-1$.
The choice of coordinates on the target space manifolds can be made through various patches. Let us choose the following patch,
the last field in the set $\{n_i\}$ (assuming it does not vanish on the selected patch) will be denoted as
\beq\label{36}
n_{N} =\varphi\,,
\eeq
where $\varphi$ will be set real. Then the coordinates on the target manifold (the Higgs branch) are
\beqn
z_i &=& \frac{ n_i }{\varphi } ,\quad i=1,2,..., N-1\,, \nonumber\\[2mm] w_a &=&{\varphi}{\rho_a},\quad a=1,2,...,{N}\,.
\label{35}
\eeqn
A useful gauge invariant parametrization is provided by the $N\times N$ matrix
\beq
M_{ia} = n_i \rho_a\,  .
\eeq
We will also introduce a radial coordinate
\beq\label{n24}
r^2 = \left( 1 +\sum_{i=1}^{N-1} \left|z_i\right|^2\right)\left(\sum_{a=1}^N \left|w_a\right|\right) \equiv {\rm Tr}\, M M^\dagger,
\eeq
cf. \eqref{M} and \eqref{r} written for the $N=2$ case.

The one-loop $\beta$ function is proportional to the Ricci tensor, while the second loop contribution is proportional to
 a convolution of the Riemann tensors,
\beq
    R^{(2)}_{p\overline{q}} = R_{p~~t}^{~rs}\riec{q}{r}{s}{t} \,.
\eeq
Both quantities were calculated in \cite{SS}. With proper coefficients inserted, the $\beta$ function takes the form 
\beq
{\beta}_{p\bar q} =  \frac{1}{2\pi} \mtr{R}{p}{q}  
	+ \frac{1}{8\pi^2} \mtr{R^{(2)}}{p}{q} + \cdots
\eeq

Furthermore, to understand how the geometry of $\mathbb{WCP}(N,N)$ model evolves, let us consider the following renormalization group (RG) equation
(valid at one loop):
\begin{equation}\label{13}
	\frac{\pd \mtr{g}{i}{j}}{\pd t} = -\frac{1}{2\pi}\mtr{R}{i}{j} \,.
\end{equation}
Here $t$ is a RG ``time", 
\begin{equation}
t=-\log\mu,
\end{equation} 
implying the larger the RG time, the lower the energy scale $\mu$ of the system.
Also, as we already mentioned on the K\"ahler manifold both the metric tensor and Ricci tensor can be expressed as double derivatives of scalar functions i.e.
\beq
	{g}_{i\bar j}= \pd_{i}\pd_{\bar{j}} K
	\quad\mbox{and}\quad
	{R}_{i\bar j} = -\pd_{i}\pd_{\bar{j}} \log{\det \mtr{g}{k}{l}}
\label{ricci}
\eeq
where $K$ is the K\"ahler potential of the manifold. Thus, we can reduce (\ref{13}) to a scalar equation
\begin{equation}\label{210}
	\frac{\pd K}{\pd t} = \frac{1}{2\pi}  \log{\det \mtr{g}{i}{j}}
\end{equation}
up to a linear combination of holomorphic and anti-holomorphic functions. 

The classical K\"ahler potential  for our theory in NLSM formulation was calculated in \cite{SVY,KS,SS,Li}. It has the 
form
\beq
K = \sqrt{\beta^{2}+4r^{2}} - \beta \log\left( \beta + \sqrt{\beta^{2} + 4r^{2}} \right) 
+ \beta \log\left( 1 + \sum_{i=1}^{N-1} |z_i|^{2} \right).
\label{classK}
\eeq
We see that it is described by the generalization of the {\em Ansatz} \eqref{ansatz} used to find the Ricci-flat conifold metric to the case of  arbitrary $N$,
\begin{equation}\label{17}
	K = f(r^{2}) + \beta \log\left( 1 + \sum_{i=1}^{N-1} |z_i|^{2} \right), 
\end{equation}
where the function $f(r^2)$ is given by
\begin{equation}\label{ic}
f^{\rm UV}(t=0,r^{2}) = \sqrt{\beta^{2}+4r^{2}} - \beta \log\left( \beta + \sqrt{\beta^{2} + 4r^{2}} \right).
\end{equation}
 The superscript ``UV'' above
shows that we will use the classical function $f(r^2)$ in Eq. \eqref{ic}  as a UV data
at $t=0$ in the RG equation. This motivates using the {\em Ansatz} \eqref{17} to describe the RG flow in NLSM
because both the UV metric and Ricci-flat conifold metric (which will be reached  in the IR) are described by the same 
{\em Ansatz} \eqref{17}. As was mentioned,
the {\em Ansatz} \eqref{17} was confirmed in perturbation theory up to two loops in \cite{SS}. 
We can also argue that {\em Ansatz} \eqref{17} is maintained by the RG flow.  Starting from  this {\em Ansatz}, at each order one convinces oneself 
that the metric determinant of such K\"ahler potential is only a function of $r^{2}$ as in Eq. (\ref{18}), so the further renormalization of the K\"ahler potential should only appear as a function of $r^{2}$. This follows from the fact that the correction to the K\"ahler potential comes from the logarithm of the metric determinant. In other words, the second term in (\ref{17}) which contributes the angular dependence to the K\"ahler potential, does not change along the RG evolution, see also \cite{PandoZayas:2000ctr}.

Let us explore the metric determinant for a generic $\mathbb{WCP}(N,N)$ model more thoroughly, namely,
\begin{equation}\label{18}
	g \equiv 	\det(\mtr{g}{i}{j})  
	=  (f')^{N-1} \left( \beta + r^{2} f' \right)^{N-1} \left( f' + r^{2} f'' \right)
\end{equation}
where $f',f''$ represent the first and second derivative with respect to $r^{2}$. 

Now, we define an auxiliary function
\begin{equation}
	\gamma(t,r^{2}) \equiv r^{2}f'(t,r^{2})\,.
	\label{gamma}
\end{equation}
cf. \eqref{23}.
We switch to $\gamma(t,r^{2})$ for the  convenience of further discussions of the RG flow, in particular, for setting the boundary condition at  each RG time. 
Then, Eq. (\ref{210}) reduces to 
\beqn
\frac{\pd f}{\pd t} &=& \frac{1}{2\pi} \Big[ (N-1)\log{f'} + (N-1)\log\big(\beta+r^{2}f'\big) \nonumber\\[2mm]
&+& \log\big(f'+r^{2}f''\big)\Big] +C\,.
\eeqn
For the time being we leave the constant $C$ undetermined. 

Before exploring the general case in the subsequent sections, let us first have a  closer
look at the simplest example, $\mathbb{WCP}(1,1)$ whose RG equation is particularly simple, namely,
\begin{equation}
	\frac{\pd f}{\pd t} = \frac{1}{2\pi} \, \log\big(f'+r^{2}f''\big) \,.
\end{equation}
A detailed analysis of this RG equation has been carried out in \cite{ARSW} where the above-formulated conjecture  on the metric flow was also confirmed. However, the results in \cite{ARSW} were formulated in the language of the original metric RG flow,  as in Eq. (\ref{13}), by virtue of
\begin{equation}
	\Omega(u) = f' + r^{2}f'' \,,
\end{equation}
where $\Omega(u)$ is the metric defined in the coordinate system $\{u,\theta\}$ used in \cite{ARSW}. Differentiating twice with respect to $u$ reproduces the metric RG equation (2.14) in \cite{ARSW} (up to a numerical factor).

\vspace{2mm}

The next in complexity case on which  we will focus for now is the RG flow in the $\wcpt$ model \footnote{Of course, we can  study the  generic $\mathbb{WCP}(N,N)$ model in the same manner, but for $N>2$  the analytic fixed point solution which is used for  comparison  below is unknown, in contradistinction with the  $\wcpt$ case.}. Let us stress that our goal in this paper is to see whether or not the ``UV" metric obtained by Higgsing $\mathbb{WCP}$ GLSM will eventually reach a Ricci-flat fixed point 
\eqref{24}, \eqref{27} in the infrared regime. Therefore, the RG equation to discuss is
\begin{align}\label{218}
	\frac{\pd f}{\pd t} = \frac{1}{2\pi} \Big[ \log{f'} + \log(\beta+r^{2}f') + \log(f'+r^{2}f'')\Big] +C,
\end{align}
with  $C$ being an integration constant. In addition, the initial condition (i.e. the function $f(r^2)$  at the UV scale) is given by \eqref{ic}.
The corresponding derivative function is 
\begin{equation}\label{114}
	\gamma^{\rm UV}(t=0,r^{2}) = \frac{2r^{2}}{\beta+\sqrt{\beta^{2}+4r^{2}}} \, .
\end{equation}
Also, a reasonable boundary condition compatible with both the initial potential and that at the IR fixed point, see \eqref{33}, is
\begin{align}\label{221}
	\gamma(t,0) =0 \,.
\end{align}
Note that (\ref{221}) is valid for any RG time rather than  just for one specific RG time.

To proceed further we note that the metric of the manifold is given by double derivatives of the
 K\"ahler potential, see \eqref{ricci}.
To take this into account we rewrite the RG equation \eqref{218} in terms of the function $\gamma$  \eqref{gamma}, which 
actually defines the metric. In particular, this allows us to get rid 
 of the undetermined integration constant $C$ in the  equation \eqref{218}.  Thus, our master RG equation takes the form
\beq
\frac{\pt \gamma}{\pt t} = \frac1{2\pi}\left\{r^2\,\frac{\gamma'^2(2\gamma +\beta) + \gamma(\gamma+\beta)\gamma''}{\gamma
(\gamma+\beta)\gamma'}-1\right\}.
\label{RGeqn}
\eeq
We will solve this equation numerically with the boundary condition \eqref{221} and initial condition \eqref{114} in the next section.

\section{Numerical solution to RG flow}
\label{numer}

In this section, we attempt to obtain the solution of (\ref{RGeqn}). In general, as most partial differential equations, it is hard to solve 
it analytically. Instead, we develop a numerical solution. 

For the purpose of solving equation (\ref{RGeqn}) we used a Runge-Kutta relaxation solver with a second order central difference discretization of the differential operators. At the borders, the method is adapted to backwards and forwards difference schemes. The spatial interval in $r^2$ is discretized on an equidistant $n_x = 100$ or $n_x=150$  point grid (depending on interval size), with a step time of order $\delta t = 10^{-3}$. A typical convergence is of the order of $\mathcal{O}(10^{3})$ iteration steps. The accuracy of the procedure is $\mathcal{O}(10^{-3})$. This accuracy can be increased by raising the number of grid points $n_x$ at the cost of iteration time. We performed several tests at higher accuracy confirming the results presented below. 

\begin{figure}[h!]
	
	\begin{subfigure}[b]{0.5\linewidth}
		\includegraphics[width=\linewidth]{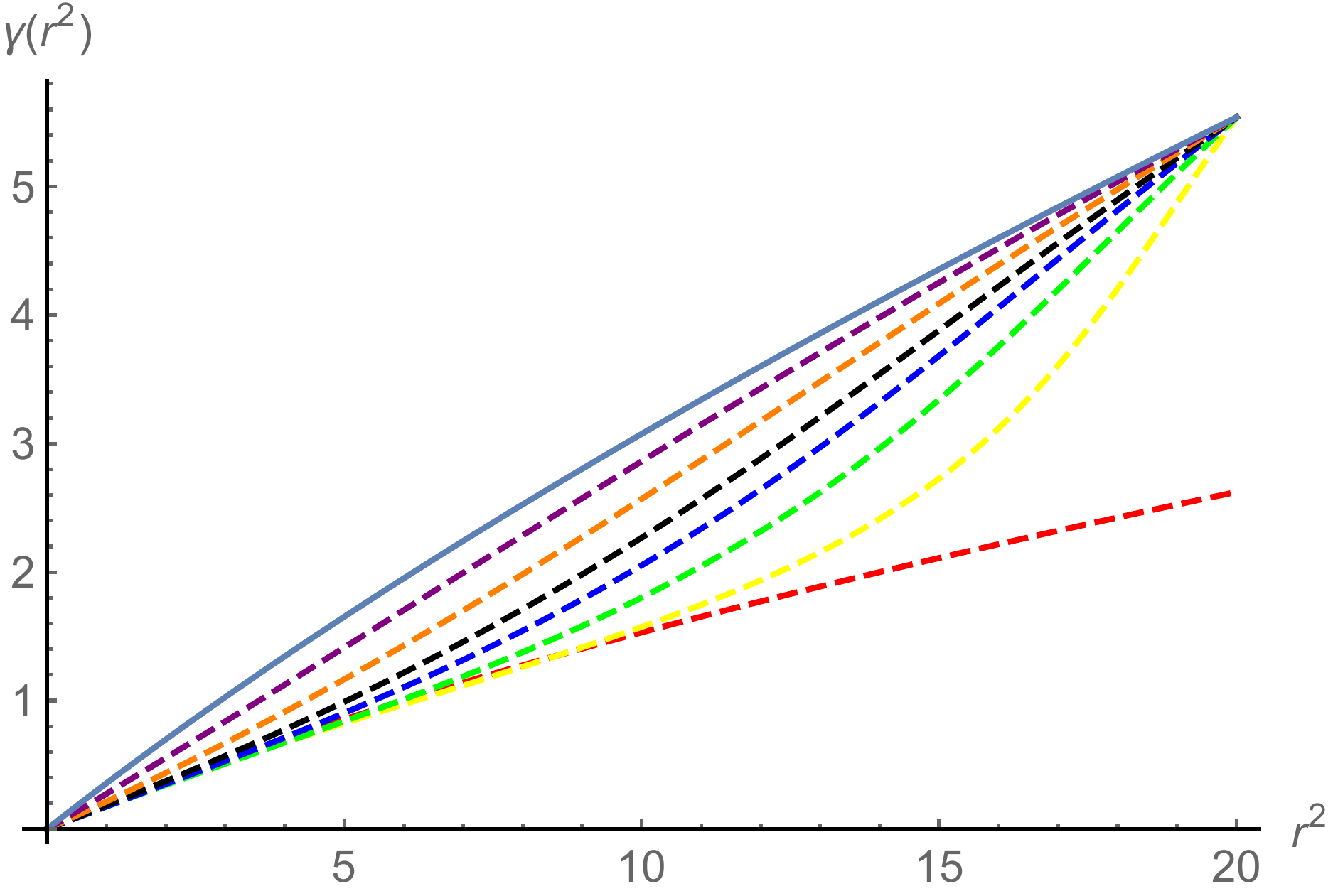}
		\caption{
$\beta=5$}
	\end{subfigure}
	\begin{subfigure}[b]{0.5\linewidth}
		\includegraphics[width=\linewidth]{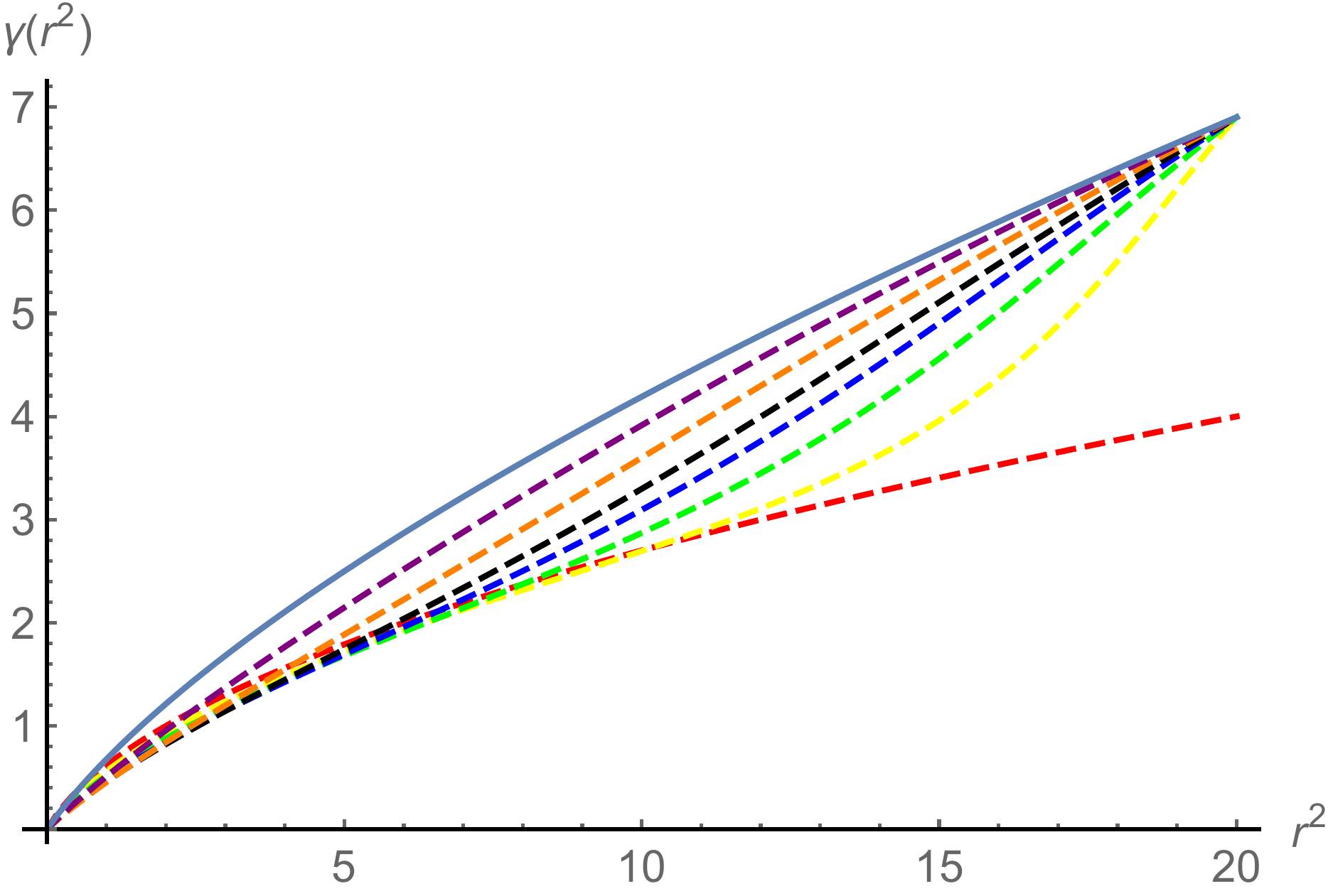}
		\caption{
$\beta=1$}
	\end{subfigure}
\center
	\begin{subfigure}[b]{0.5\linewidth}
\center
		\includegraphics[width=\linewidth]{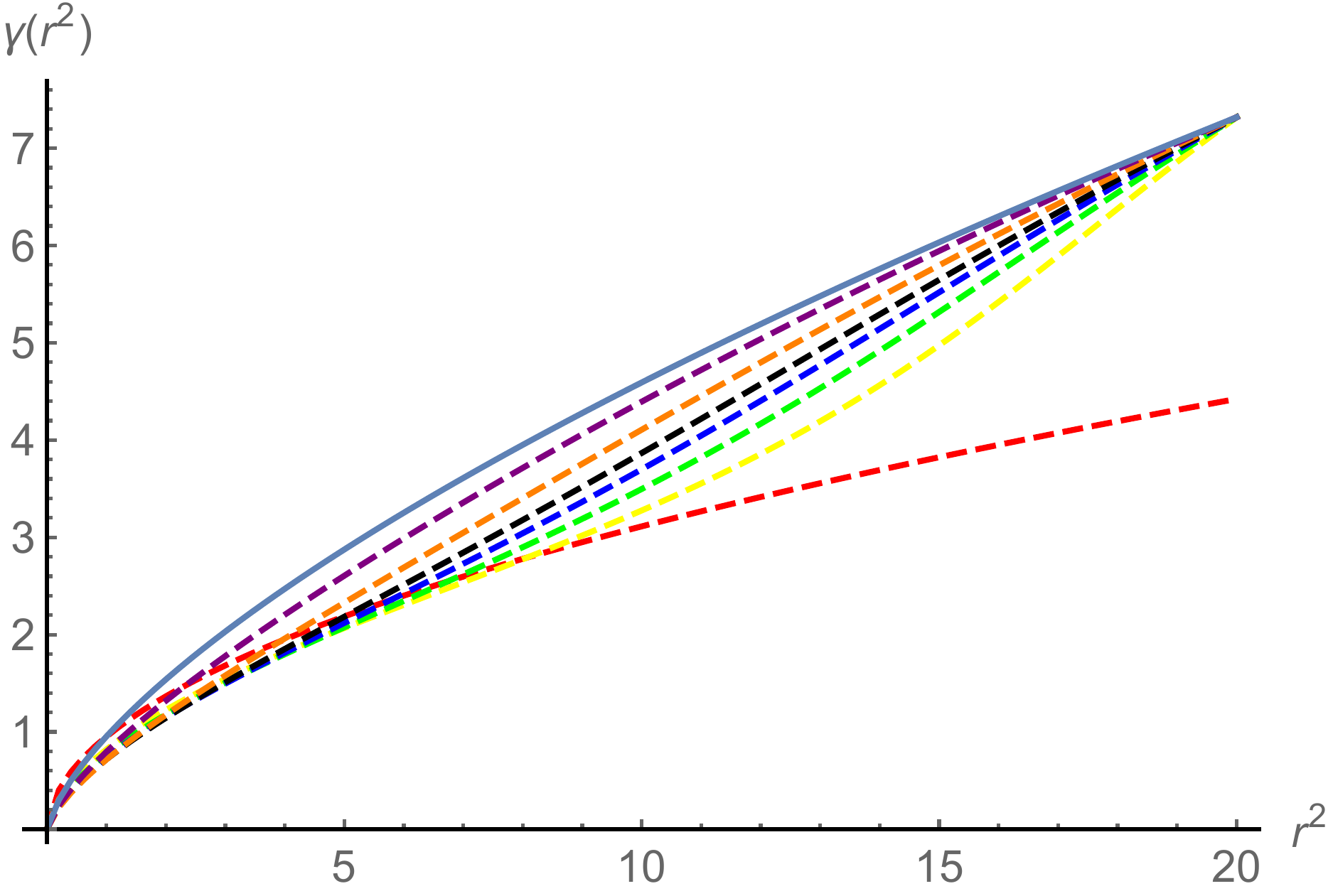}
		\caption{
$\beta=0.1$}
	\end{subfigure}

	\caption{\small Convergence of the IR flow of $\gamma(r^2)$ from the starting UV point $\gamma = 2r^2/\sqrt{\beta^2+4 r^2}$ (red dashed line) to the analytic IR solution (solid blue line). From the bottom to the top dashed lines (yellow to purple) represent the intermediate K\"ahler potentials from the early RG time to the late RG time. Also, in this case, the boundary condition for iteration flows is set to match the analytic solution at $r^{2}=20$.}
	\label{fig:conv1}
\end{figure}

\begin{figure}[h!]
	
	\begin{subfigure}[b]{0.5\linewidth}
		\includegraphics[width=\linewidth]{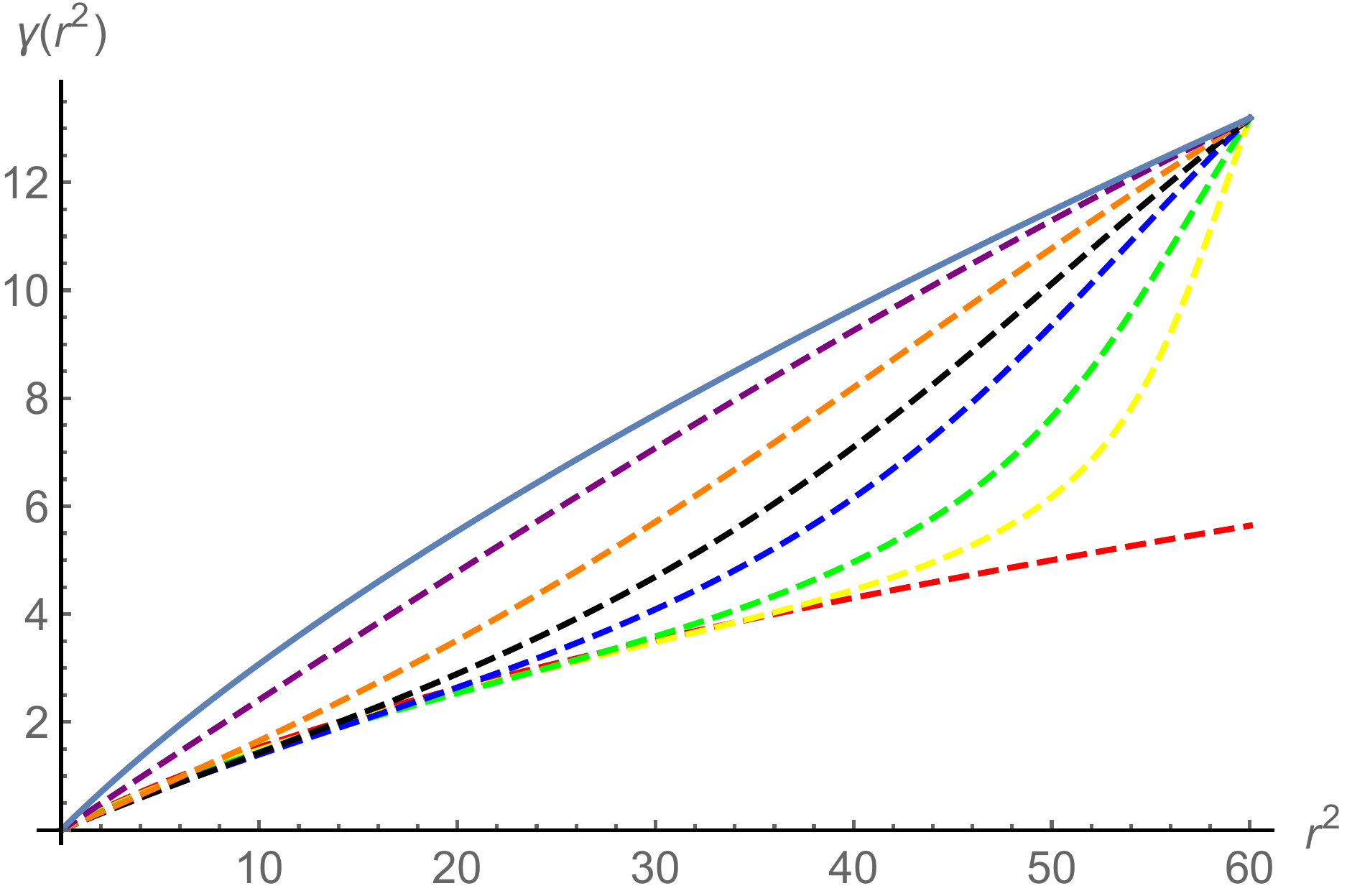}
		\caption{
$\beta=5$}
	\end{subfigure}
	\begin{subfigure}[b]{0.5\linewidth}
		\includegraphics[width=\linewidth]{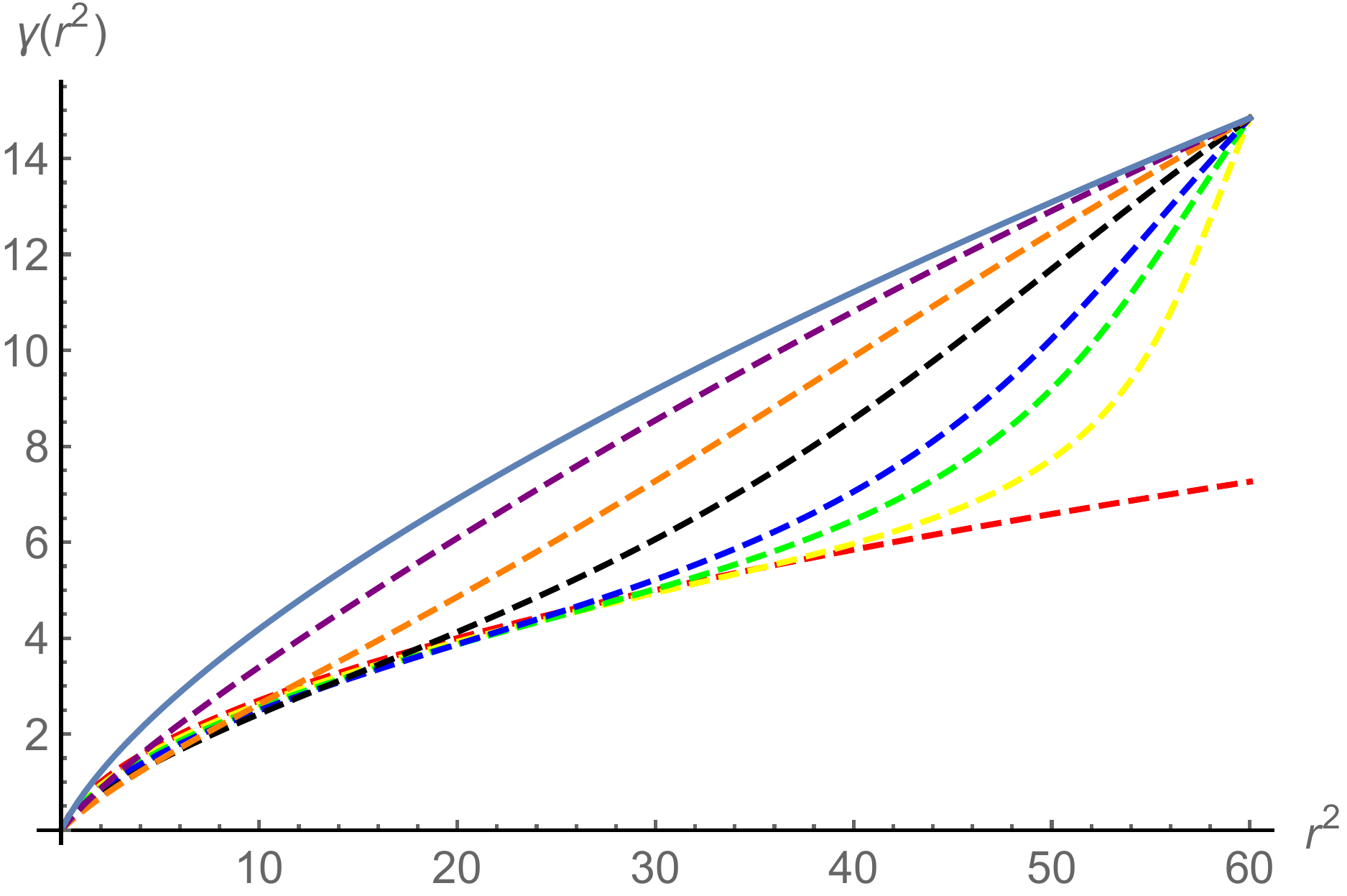}
		\caption{
$\beta=1$}
	\end{subfigure}
\center
	\begin{subfigure}[b]{0.5\linewidth}
\center
		\includegraphics[width=\linewidth]{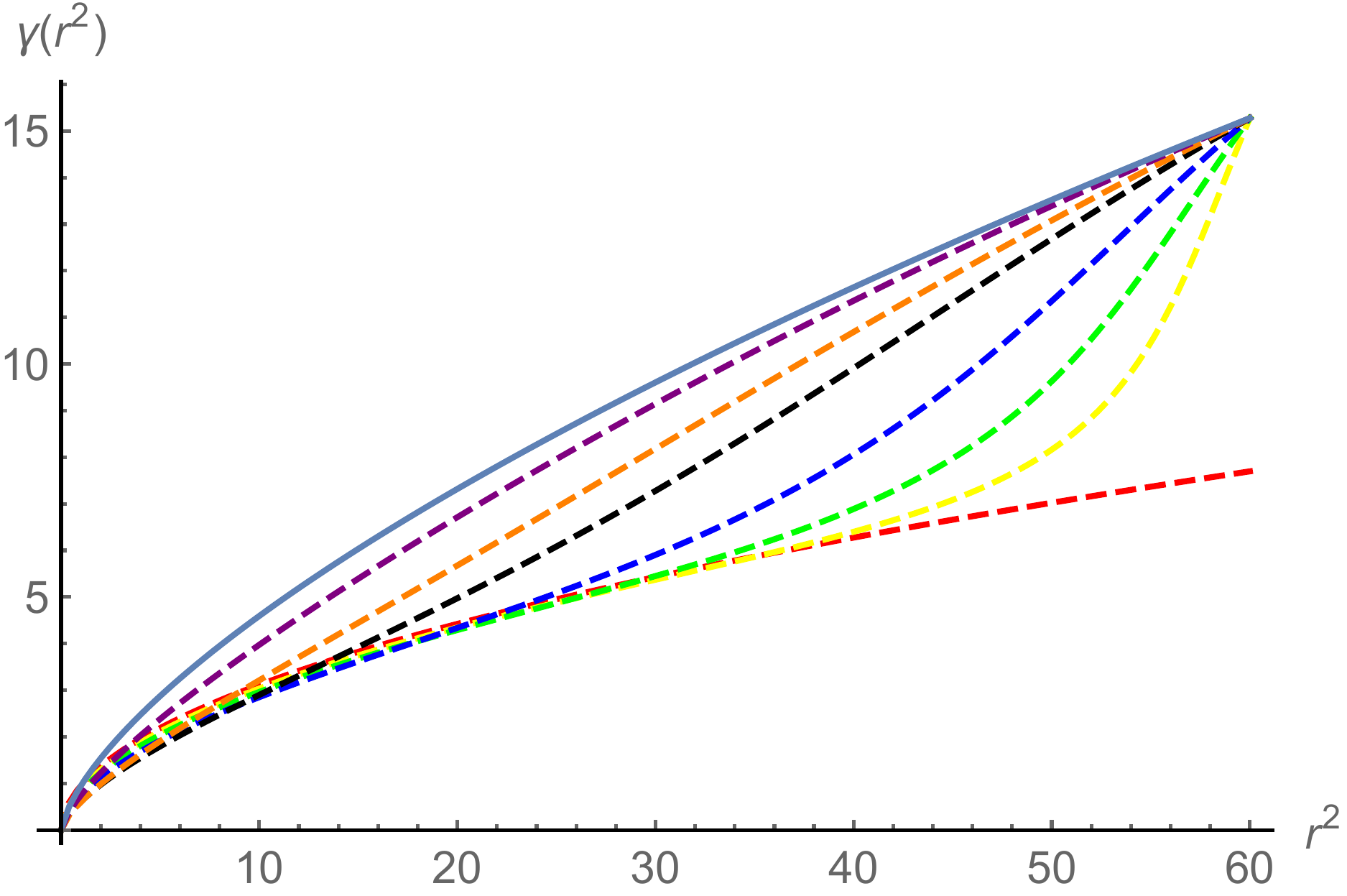}
		\caption{
$\beta=0.1$}
	\end{subfigure}

	\caption{\small Larger $r^2$ test of convergence of the IR flow of $\gamma(r^2)$ from the starting UV point $\gamma = 2r^2/\sqrt{\beta^2+4 r^2}$ (red dashed line) to the analytic IR solution (solid blue line). From the bottom to the top dashed lines (yellow to purple), they represent the intermediate K\"ahler potentials from the early RG time to the late RG time. Also, in this case the boundary condition for iteration flows is set to match the analytic solution at $r^{2}=60$.}
	\label{fig:conv2}
\end{figure}

In our numerical solution we set as boundary condition that $\gamma(L)=\gamma_{IR}(L)$, where $L$ is the size of the $r^2$ interval and $\gamma_{IR}$ is given by equation (\ref{24}). In order to show this is a well-defined boundary condition, two signals are presented in the following order. First, the notion of large $r^{2}$ is relative to $\beta$ since the only scale in the theory is $\beta$. Then, for different boundary points set at any sufficiently large $r^{2}$, the flows all converge to the fixed point solution (\ref{24}) as shown in Fig. \ref{fig:conv1}. Furthermore, we tested the convergence by changing the size of the $r^2$ interval, finding convergence again as shown in Fig. \ref{fig:conv2}.

The converging curves in the above two graphs also show that the larger $r^{2}$ we fix, the longer time they would take to converge to the solution at the fixed point, which indicates that once such fixing condition is performed at $r^{2}=\infty$, the curve actually take infinite amount of RG time to converge.
The other evidence for the convergence of this numerical calculation is that if the boundary point is fixed at the same large $r^{2}$ for different $\beta$, the flow stably converges to the corresponding fixed point solution as the smaller time step is applied.

\section{Exact twisted superpotential of $\mathbb{WCP}\boldmath{(N,N)}$}	
\label{sec51}
\setcounter{equation}{0}

In the previous section we use numerical methods to study the RG flow of the K\"ahler potential in the $\mathbb{WCP}(2, 2)$ model. On the other hand, we also want to study the vacuum structure of $\mathbb{WCP}(N, N)$ model. If the theory in deep IR flows to a conformal fixed point, there would be no dynamical mass generated. This is  in contrast to $\mathbb{CP}(N-1)$ model 
\cite{W79} where a dynamical mass is generated due to a VEV of  $|\sigma|^2$ in the vector supermultiplet, see in Eq.\,(\ref{cpg}). In this section, 
we use the exact twisted superpotential of the theory \cite{AdDVecSal,W2} to find the  $\sigma$ VEV.


To write it down for the case $\tN=N$ we introduce  two sets of twisted masses $\{\tilde m_k\}$ and $\{\hat m_a\}$ for chiral matters $n_k$ and $\rho_a$, respectively, \cite{HaHo, DHT} as an IR regularization. In the end we put all masses to zero.
Upon integrating out all matter fields it takes the form
\beqn
\mathcal W_{\rm eff}(\Sigma)\=-\frac{\beta_h}{2} \sqrt{2}\Sigma-\frac{1}{4\pi}\sum_{k=1}^N(\sqrt{2}\Sigma+\tilde m_k)
\left[\log\left(\frac{\sqrt{2}\Sigma+\tilde m_k}{\mu}\right)-1\right]\nonumber\\
\+\frac{1}{4\pi}\sum_{a=1}^N(\sqrt{2}\Sigma+\hat m_a)\left[\log\left(\frac{\sqrt{2}\Sigma+\hat m_a}{\mu}\right)-1\right]\,.
\eeqn
Here $\beta_h$ is the complexified FI-coupling, 
\beq
\beta_h = \beta +i\frac{\theta}{2\pi}\,,
\label{k62}
\eeq
 where $\theta$ is the $\theta$ angle.

The VEV of $\sigma$, to be denoted as  $\Sigma$, is therefore given by the solution to 
\beq
\frac{\partial\mathcal W_{\rm eff}}{\partial\Sigma}\Bigg{|}_{\Sigma=\langle \sigma \rangle}=0\,,
\eeq
i.e.
\beq
\prod_{k=1}^N\left(\langle \sqrt{2}\sigma \rangle+\tilde m_k\right)=e^{-2\pi  \beta_h}\prod_{a=1}^N\left(\langle \sqrt{2}\sigma \rangle+\hat m_a\right)\,.
\eeq
When the twisted masses $\{\tilde m_k\}$ and $\{\hat m_a\}$ tend to zero, the only solution to the above equation is
\beq
\langle \sigma \rangle=0\,,
\eeq
for any nonvanishing  $\beta_h$. The zero value for $\sigma$ means that the mass gap for $n$ and $\rho$ fields is not generated and we are in the conformal regime. Note, that if $\beta =0$ there is another solution with an arbitrary non-zero $\sigma$. 
This solution describes the Coulomb branch which opens up at $\beta =0$. As was already mentioned, we do not consider 
the case of vanishing $\beta$ in this paper.

\section{Renormalization in GLSM vs. NLSM}	
\label{sec4}
\setcounter{equation}{0}

Another main observation in \cite{SS} is that, even though the FI-coupling constant $\beta$ receives no quantum corrections in the GLSM/NLSM, the NLSM K\"ahler potential still has logarithmic divergences and thus evolves with the energy scale $\mu$ or RG-time $t$. It has been mentioned in Sec.\ref{intro}, see also in \cite{ARSW}, that the  $n_i$ and $\rho_a$ 
$Z$-factors are not protected and therefore the RG flow is not limited to one loop. The non-trivial quantum corrections to the K\"ahler potential in NLSM are due to these $Z$ factors. In this section, we will make the statement concrete.

In fact, to understand this phenomenon from the GLSM viewpoint, it would be more appropriate to study the anomalous dimensions of the meson operators
\beq
M_{ia}=n_i\rho_a\,,\ \ \ {\rm and}\ \ \ Z_i^j=\frac{n_j}{n_i}\,.
\label{k71}
\eeq
Indeed, the above gauge invariant moduli span the vacuum manifold (the Higgs branch of GLSM). The running of the K\"ahler potential in NLSM is due to renormalization of the classically marginal mesons operators (\ref{k71}). In the GLSM formalism, their runnings is described by their anomalous dimensions and can be computed perturbatively,
\beq
\gamma_{M_{ia}}=-\mu\frac{\partial}{\partial\mu}Z_{M_{ia}},\ \ \ {\rm and}\ \ \ \gamma_{Z_{i}^j}=-\mu\frac{\partial}{\partial\mu}Z_{Z_{i}^j}\,.
\eeq

To calculate the anomalous dimensions of $M_{ia}$ and $Z_{i}^{j}$, it is convenient to invoke the superfield formulation of GLSM. Equation (\ref{cpg}) can be obtained from the Lagrangian
\begin{align}\label{sqed}
	{\cal L} = {\cal L}_{\rm v.m} 
	+ \int d^{4}{\theta} \, \Big(
	\bar{N}_{i} e^{V} N_{i} + \bar{R}_{a} e^{-V} R_{a} - \beta V
	\Big)
\end{align}
where ${\cal L}_{\rm v.m}$ is the Lagrangian of the vector multiplet. Note that $N_{i}$ and $R_{a}$ are the chiral multiplets with charges $1$ and $-1$, respectively, and the summation of the indices $i,a=1,2,\cdots,N$ is performed. 

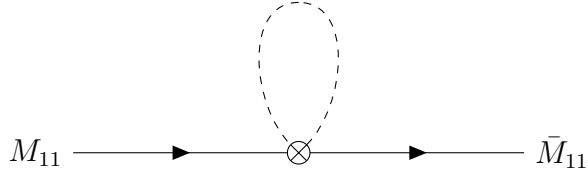
\begin{figure}[h]
	\begin{align*}
	\begin{tikzpicture}
	\begin{feynman}
	\vertex (b0){\(M_{11}\)};
	\vertex[right=3.5cm of b0,crossed dot] (b1) {};
	\vertex[right=3.5cm of b1] (b2){\(\bar{M}_{11}\)};
	\vertex[above=2 cm of b1] (b3);
	\diagram*{
		(b0) --[fermion](b1) --[fermion](b2),
		(b1) --[scalar,out=45, in=0,min distance=0.5cm] (b3) --[scalar,out=180,in=135,min distance=0.5cm](b1),
	};
	\end{feynman}
	\end{tikzpicture}
	\end{align*}
	\vspace{-5ex}
	\caption{\small The crossed dot indicates the vertices essentially originates from the contraction of $V$ superfield (see below for detailed explanation) and the explicit expression can be read off from (\ref{pertb}). The dashed line denote both moduli, $Z_{1}^{j}$ and $M_{1a}$ proparating in the loop.}
	\label{tad4}
\end{figure}

On the Higgs branch, the chiral multiplets develop  VEVs; for simplicity let us choose
\begin{align}
	\left| N_{1} \right|^{2} = \beta \,.
	\label{k74}
\end{align}
Then, (\ref{sqed}) can be recast in terms of the moduli, namely,
\begin{align}
	\int d^{4}{\theta} \, \Bigg\{
	\beta e^{V} \left( 1 + \sum_{j=2}^{N} \left| Z_{1}^{j} \right|^{2} \right) 
	+ \frac{1}{\beta} e^{-V} \sum_{a=1}^{N}\left|M_{1a} \right|^{2} - \beta V
	\Bigg\} \,.
\end{align}
To study the Z-factor correction under this particular vacuum (\ref{k74}) we eliminate the massive $V$ superfield by solving the equation of motion for the vector multiplet,
\begin{align}
	e^{V_0} = \frac{
	\beta + \left[ \beta^{2} + 4 \left( 1 + \sum_{j=2}^{N} \left| Z_{1}^{j} \right|^{2} \right) \left( \sum_{a=1}^{N} \left| M_{1a} \right|^{2} \right) \right]^{1/2}
	}{2\beta\left( 1 + \sum_{j=2}^{N} \left| Z_{1}^{j} \right|^{2} \right)}\,.
\label{k76}	
\end{align} 
Here $V_0$ is the classical solution.
In the weak coupling limit $\beta = \frac{2}{g^{2}} \gg 1$, upon substitution of (\ref{k76}), Eq. (\ref{sqed}) takes the form
\begin{align}\label{pertb}
	{\cal L} = {\cal L}_{\rm v.m} +&
	\int d^{4}{\theta}\, \Bigg\{
	\frac{1}{\beta} \sum_{a=1}^{N}\left|M_{1a}\right|^{2}  
	+\beta\sum_{j=2}^{N}\left|Z_{1}^{j}\right|^{2}
	-\frac{1}{2\beta^{3}} \sum_{a,b=1}^{N} \left| M_{1a}\right|^{2}\left| M_{1b}\right|^{2}
	\nonumber\\[2mm]
	&
	+\frac{1}{\beta}\sum_{a=1,j=2}^{N} \left|Z_{1}^{j}\right|^{2}\left|M_{1a}\right|^{2}
	-\frac{\beta}{2} \sum_{j,k=2}^{N} \left| Z_{1}^{j}\right|^{2}\left| Z_{1}^{k}\right|^{2}
	+{\cal O}(\delta V^{2})
	\Bigg\}
\end{align}
Since the overall structure of the K\"ahler potential here is manifest, let us trace only the renormalization of the term $M_{11}\bar{M}_{11}$. The logarithmic one-loop correction results from the tadpole graphs emerging from four-M terms in (\ref{pertb}) as shown in Fig. \ref{tad4},
\begin{align}\label{725}
	-\frac{1}{2\pi\beta^2}\log\frac{M_{V}}{\mu} \cdot 
	\left[  \frac{2 \cdot 2}{2} +  \frac{2 (N-1)}{2} - (N-1) \right]
	=-\frac{1}{\pi\beta^2}\log\frac{M_{V}}{\mu}
\end{align}
where the first term in the square bracket comes from $\large(\bar M_{11} M_{11}\large)^2$ while the latter two terms come from the mixed  terms $\large(\bar M_{11} M_{11}\large) \large(
\bar M_{1a} M_{1a}\large )$
and a similar one with  $M_{1a}$ and $Z_{1}^{j}$ moduli. Assembling all contributions we arrive at the coefficient in front of  $\bar{M}_{11} M_{11}$ 
\begin{align}
	\frac{1}{\beta} M_{11}\bar{M}_{11} \to 
	\frac{1}{\beta} \left( 1- \frac{1}{\pi\beta}\log\frac{M_{V}}{\mu} \right)M_{11}\bar{M}_{11} \,.
\end{align}
The $N$-independence of the result is explained by the fact of a dis-balance of the $n$ fields (one of $N_{i}$ develops a VEV). 

Before demonstrating that the above result coincides with that in the NLSM formalism, it is instructive to verify our previous claim that $Z_{i}^{j}$ has no Z-factor correction from the direct computation. That is, from the tadpole diagrams similar to those in Fig. \ref{tad4}, we see that the correction to $\bar{Z}_{1}^{j}Z_{1}^{j}$ is
\begin{align}
	\delta Z_{\bar{Z}_{1}^{j}Z_{1}^{j}}
	= -\frac{1}{2\pi\beta}\log\frac{M_{V}}{\mu} \cdot
	\left[ \frac{2 \cdot 2}{2} +  \frac{2 (N-2)}{2} - N \right]
	=0\,.
\end{align}
The multiplicity of tadpoles producing the logarithmic divergences is counted in the same way and order as presented in (\ref{725}):  i.e. $Z_{1}^{j}Z_{1}^{j},\,\, Z_{1}^{j}Z_{1}^{k}$ and $Z_{1}^{j}M_{1a}$.

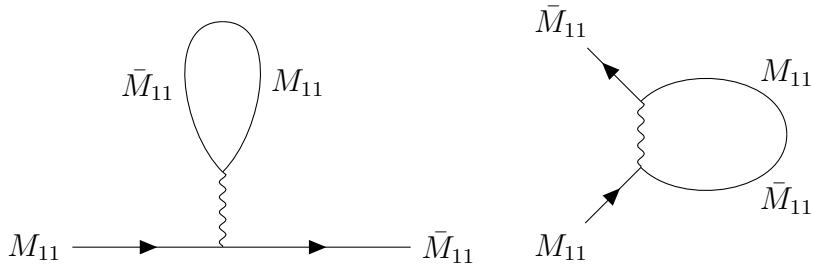
\begin{figure}[h]
	\begin{align*}
	\begin{tikzpicture}
	\begin{feynman}
	\vertex (b0){\(M_{11}\)};
	\vertex[right=2.5cm of b0] (b1);
	\vertex[right=2.5cm of b1] (b2){\(\bar{M}_{11}\)};
	\vertex[above=1 cm of b1] (b3);
	\vertex[above=2 cm of b3] (b4);
	\vertex[below right=.7 cm of b4] (b5) {\(M_{11}\)};
	\vertex[below left=.7 cm of b4] (b6) {\(\bar{M}_{11}\)};
	\diagram*{
		(b0) --[fermion](b1) --[fermion](b2),
		(b1) --[photon] (b3),
		(b3) --[out=45, in=0,min distance=0.5cm] (b4) --[out=180,in=135,min distance=0.5cm](b3),
	};
	\vertex[right=3cm of b2] (a2);
	\vertex [left=1cm of a2](a1) {\(M_{11}\)};
	\vertex [above=3 cm of a1](a3){\(\bar{M}_{11}\)};
	\vertex [above right=1.5cm of a1](a4);
	\vertex [below right=1.5cm of a3](a5);	
	\vertex[above=1.5 cm of a1] (a6);
	\vertex[right=3cm of a6](a7);
	\vertex[above=.5cm of a7](a8) {\(M_{11}\)};
	\vertex[below=.5cm of a7](a9) {\(\bar{M}_{11}\)};
	\diagram*{
		(a3) --[anti fermion](a5),
		(a1) --[fermion](a4),
		(a4) --[photon](a5),
		(a4) --[out=-45, in=-90,min distance=0.5cm](a7)
		--[out=90, in=45,min distance=0.5cm](a5),
	};
	\end{feynman}
	\end{tikzpicture}
	\end{align*}\vspace{-5ex}
	\caption{\small The solid line represents the propagator of the light fields (or the corresponding moduli) and the curvy line is $V$ as it emerges from a blow up of the  effective vertex (crossed circle) in Fig. \ref{tad4}.} 
	\label{tadV}
\end{figure}

As a concluding remark, let us explore how the heavy vector multiplet produces the effective vertices. For simplicity, let us focus on the case of $\mathbb{WCP}(2,2)$ and examine  only the term $\bar{M}_{11}M_{11}$. Assume on the Higgs branch the field $N_{1}$ acquires a VEV and is thought of as a ``heavy" field. We then have three light fields which can produce logarithms in the tadpole loop, namely, $N_{2}$ and $R_{1,2}$. The four-leg operators comprising $M_{11}$ can be established via $-\bar{R}_{1,2}R_{1,2}V$ and $\bar{N}_{2}N_{2}V$ vertices upon contraction of the superfield $V$, and, indeed, 
\begin{align*}
	(\bar{R}_{1}R_{1})^{2} + 2\bar{R}_{1}R_{1}\left[ \bar{R}_{2}R_{2}
	- \bar{N}_{2}N_{2}\right]
	\sim
	\left|M_{11}\right|^{4} 
	+ 2\left|M_{11}\right|^{2} \left[ \left|M_{12}\right|^{2}-\beta^{2} \left|Z_{1}^{2}\right|^{2} \right] \,
\end{align*}
as the latter directly emerges from the expansion of the $\mathbb{WCP}$ Lagrangian. In particular, if we want to restore $V$ and see how $\bar{M}_{11}M_{11}$ is convoluted, the tadpole diagram in Fig. \ref{tad4} can be viewed as the combination of the two graphs in Fig. \ref{tadV}.

\section{Comparison with the \boldmath{$zn$} model}
\label{zn}
\setcounter{equation}{0}

In the previous sections, we analyze the UV to IR flow of the  $\mathbb{WCP}(2,2)$ model from different perspectives. 
In fact, it is not this particular model which emerges on the world sheet of an appropriate semilocal string. The so-called $zn$ model, which is close but not quite identical to $\mathbb{WCP}(2,2)$  
emerges \cite{SVY,KS}. 
In this section we will study it following the same line of reasoning.

To begin with, recall that the $zn$-model consists of two kinds of massless complex fields
$\tilde{n}_{k}$ and $\tilde{z}_{a}$ for $k,a=1,\ldots,N$, and a $U(1)$ gauge field $A_{\mu}$. The action of $zn$ model in gauge formulation reads
\begin{align}\label{71}
	S_{zn} = \int d^{2}x\, \Big\{
	 \left|\nabla_{\mu} \tilde{n}_{k}\right|^2 
	 &+  \left|\pt_\mu(\tilde{n}_{k}\tilde{z}_{a})\right|^2  + \frac1{4e^2}F^2_{\mu\nu} + \frac1{e^2}
	|\pt_\mu\sigma|^2+\frac1{2e^2}D^2
	 \nonumber\\[2mm]
	&+    2|\sigma|^2  |\tilde{n}_{k}|^2 + iD \left(|\tilde{n}_{k}|^2  -\beta\right)
	\Big\}+\mbox{fermions}.
\end{align}
Note that the $U(1)$ gauge field $A_{\mu}$ acts on $\tilde{n}_{k}$ through the covariant derivative $\nabla_{\mu}$ as defined in (\ref{22}), while
the  $\tilde{n}_{k}\tilde{z}_{a}$ operator (i.e. the second term in (\ref{71})) is neutral.

On the Higgs branch and after integrating out the heavy gauge field we arrive at the theory whose K\"ahler potential is \cite{SVY}
\begin{align}\label{72}
	K^{{\rm UV}}_{zn} = \left|\zeta\right|^{2} + \beta\log(1+\left|\Phi_{j}\right|^{2})
\end{align}
where
\begin{align}
\left|\zeta\right|^{2} = \left|{\cal Z}_{a}\right|^{2}(1+\left|\Phi_{j}\right|^{2}), \quad\mbox{with}\quad
{\cal Z}_{a} = \tilde{z}_{a}\tilde{n}_{N}, \quad \Phi_{j} = \frac{\tilde{n}_{j}}{\tilde{n}_{N}} \,,
\end{align}
$a=1,\ldots,N$ and $j=1,\ldots,N-1$. Here we choose a coordinate patch with $\tilde{n}_{N}$ non-vanishing. $\left|\zeta\right|^{2}$ is an invariant radial coordinate playing the same role as $r^{2}$ in the previous $\mathbb{WCP}$ model. We stress that (\ref{72}) has the same type of K\"ahler potential as (\ref{17}) and therefore the formulae developed in Sec. \ref{geoboson} can be directly applied. In particular, the Ricci tensor that determines the renormalization of the K\"ahler potential 
is given by the second equation in \eqref{ricci}, where the determinant of the metric is given by
\begin{align}
	\det(g)_{zn}^{{\rm UV}} = (\beta+\left|\zeta\right|^{2})^{N-1}.
	\label{gdet}
\end{align}

\subsection{Z factors}

 Let us  examine  the one-loop correction to the K\"ahler potential. The metric determinant is given by \eqref{gdet}
leading to the following correction in the K\"ahler potential:
\begin{align}\label{75}
	\Delta K_{zn}^{(1)} = \frac{1}{2\pi} \log\frac{M_{V}}{\mu} \cdot (N-1)\log(\beta+\left|\zeta\right|^{2}) \,
\end{align}
which coincides with the one given in \cite{KS}. The former logarithm comes from the loop integral while the latter factor $(N-1)\log(\beta+\left|\zeta\right|^{2})$ originates from the metric determinant. Because the FI term $\beta$ here  does not run in the RG process, the correction (\ref{75}) cannot be attributed to it. In fact, this additional logarithm is associated with the similar Z factor as was discussed\,\footnote{The contribution of the $Z$ factor from another gauge invariant parameter $\tilde{n}_{i}/\tilde{n}_{N}$ vanishes by the same reason mentioned in Sec. \ref{sec4} for $Z^{a}_{b}$.} in Sec. \ref{sec4}.
First, at ${\cal O}(\beta^{-1})$ level, the correction to the K\"ahler potential gives
\begin{align}\label{76}
\Delta K_{zn}^{(1)} 
\approx \frac{N-1}{2\pi\beta} \log\frac{M_{V}}{\mu} \cdot \left|\zeta\right|^{2} \,,
\end{align}
in the vicinity of the origin. Note that $\left|\zeta\right|^{2}$ is similar to $r^{2}$ in the $\mathbb{WCP}(2,2)$ model which can also be expressed as
\begin{align}
	\left|\zeta\right|^{2} = {\rm Tr}\tilde{M}\tilde{M}^{\dagger}
	\quad\mbox{with}\quad
	\tilde{M}_{ka} \equiv \tilde{n_{k}}\tilde{z}_{a} \,.
\end{align}
On the GLSM side, to calculate the $Z$ factor of meson operator, it suffices to calculate the $Z$ factor of $\tilde{z}$ field. To see this is the case, we can look at the second term in (\ref{71}). This term is like a meson kinetic term 
\begin{align}\label{mk}
	\left|\pt_\mu\tilde{M}_{ka}\right|^2
	 = \left|\tilde{n}_k\right|^{2}\pt \tilde{z}_a\pt\overline{\tilde{z}_a} + 
	 \left|\tilde{z}_a\right|^{2}\pt \tilde{n}_k\pt\overline{\tilde{n}_k} + 
	 \tilde{n}_k\overline{\tilde{z}_a}\pt\tilde{z}_k\pt\overline{\tilde{n}_a}+
	 \tilde{z}_a\overline{\tilde{n}_k}\pt\tilde{n}_k\pt\overline{\tilde{z}_a} \,.
\end{align}
\begin{figure}[t]
	\begin{align*}
	\begin{tikzpicture}
	\begin{feynman}
		\vertex (b0){\(\pt\tilde{z}_{a}^{0}\)};
		\vertex[right=2cm of b0] (b1);
		\vertex[right=2cm of b1] (b2){\(\pt\overline{\tilde{z}_{a}^{0}}\)};
		\vertex[above=1.5 cm of b1] (b3);
		\vertex[below=1.2cm of b1] (b4) {(1)};
		\diagram*{
			(b0) --[fermion](b1) --[fermion](b2),
			(b1) --[scalar, out=45, in=0,min distance=0.5cm] (b3) --[scalar,out=180,in=135,min distance=0.5cm](b1),
		};
		\vertex[right=3cm of b2] (a2);
		\vertex [left=1cm of a2](a1) {\(\pt\tilde{z}_{a}^{0}\)};
		\vertex [right=3cm of a2](a3);
		\vertex [right=1 cm of a3](a4){\(\pt\overline{\tilde{z}_{a}^{0}}\)};	
		\vertex[right=7cm of b4] (a5) {(2)};	
		\diagram*{
			(a2) --[anti fermion](a1),
			(a2) --[anti charged scalar,out=45,in=135,edge label=\(l\)] (a3) --[fermion,edge label=\(p\)] (a4),
			(a2) --[charged boson,out=315,in=225] (a3),
		};
	\end{feynman}
	\end{tikzpicture}
	\end{align*}
	\caption{\small The solid lines are the background field $\pt\tilde{z}_{a}^{0}$ and $\pt\overline{\tilde{z}_{a}^{0}}$ while the dashed lines present $\tilde{n}_{k}^{q}$ and the curvy 
	line is the quantum $\tilde{z}$ field, $\tilde{z}^{q}_{a}$. In the second diagram, the dashed line propagator is only for $\tilde{n}_{N}^{q}$ as indicated in (\ref{710}).}
	\label{zzn}
\end{figure}
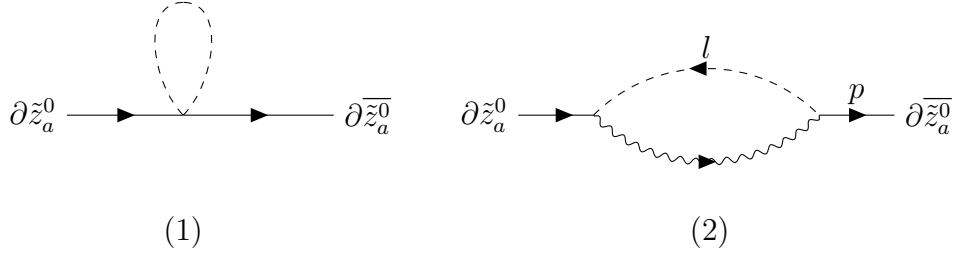\\
Once $\tilde{M}_{ka}$ obtains a $Z$-factor renormalization, it will also be reflected on the right hand side of (\ref{mk}) and {\em vice versa}. In particular, from the first term in (\ref{mk}), we see that this is indeed a wave-function renormalization of $\tilde{z}$ field if we expand around the vacuum,
\begin{align}
	\tilde{n}_{j}^0=0,\quad
	\tilde{n}_{N}^0 = \sqrt{\beta},\quad
	\tilde{z}^{0}(x)
	\quad\mbox{for}~~j=1,\ldots,N-1\,.
\end{align}
In  this background, (\ref{mk}) takes the form
\begin{align}\label{710}
	  \left|\pt_{\mu} \tilde{n}_{k}\right|^2+\left|\pt_\mu\tilde{M}_{ka}\right|^2
	 &= \beta \, \pt \tilde{z}_a^{0}\pt\overline{\tilde{z}_a^0} 
	 + \beta \, \pt \tilde{z}_a^{q}\pt\overline{\tilde{z}_a^q} 
	 + \pt \tilde{n}_k^{q}\pt\overline{\tilde{n}_k^q} 
	 + \tilde{n}_k^{q}\overline{\tilde{n}_k^q} \pt \tilde{z}_a^{0}\pt\overline{\tilde{z}_a^0} \nonumber\\[2mm]
	 &+\tilde{n}_k^{q}\overline{\tilde{n}_k^q}\tilde{z}_a^{q}\overline{\tilde{z}_a^q} 
	 +\tilde{n}_{N}^0\overline{\tilde{z}^{q}_{a}}\pt\tilde{z}_{a}^0\pt\overline{\tilde{n}_{N}^{q}}
	 +\overline{\tilde{n}_{N}^0}\tilde{z}^{q}_{a}\pt\overline{\tilde{z}_{a}^0}\pt\tilde{n}_{N}^{q}
	 +\cdots\,,
\end{align}
where $\tilde{n}_{k}^{q}$ and $\tilde{z}_{a}^{q}$ are the quantum parts of $\tilde{n}_{k}$ and $\tilde{z}_{a}$ fields, respectively. Here we only list the propagators and vertices we will use. The first diagram in Fig. \ref{zzn} contributes
\begin{align}
	N \cdot i\int\,\frac{d^{2}l}{(2\pi)^{2}} \frac{1}{l^2} = \frac{N}{2\pi}\log\frac{M_V}{\mu} \,,
\end{align}
while the contribution from the second diagram is
\begin{align}
	\left|\tilde{n}^0_N\right|^2 \, \pt \tilde{z}_a^{0}\pt\overline{\tilde{z}_a^0} \cdot
	(-i)\int\frac{d^{2}l}{(2\pi)^{2}} \frac{l^2}{l^2 \cdot \beta(l+p)^2}
	= -\frac{1}{2\pi}\log\frac{M_{V}}{\mu} \cdot \pt \tilde{z}_a^{0}\pt\overline{\tilde{z}_a^0} \,,
\end{align}
where we plug in the background field $\left|\tilde{n}^0_{N}\right|^2=\beta$. 
Combing the above two graphs, we thus find  the $Z$-factor of the $z$ fields at one-loop
\begin{align}
	\beta \, \pt \tilde{z}_a^{0}\pt\overline{\tilde{z}_a^0} 
	\to
	\left( 1 + \frac{N-1}{2\pi\beta}\log\frac{M_V}{\mu} \right) \cdot \beta \, \pt \tilde{z}_a^{0}\pt\overline{\tilde{z}_a^0} \,.
\end{align}
This indeed matches the additional logarithm shown in the NLSM one-loop effective K\"ahler potential (\ref{76}). Note that we can also expand around another point on the vacuum manifold
\begin{align}
	\sum_{k=1}^{N} \left|\tilde{n}_{k}\right|^2 = \beta \,.
\end{align}
Nothing changes for the first diagram while in the second one, there would be $N$ replicas with all the same logarithmic contribution and the coefficient $\left|\tilde{n}_{k}^{0}\right|^2$ (no sum). The overall coefficient in front of $\pt \tilde{z}_a^{0}\pt\overline{\tilde{z}_a^0}$ becomes $\sum\left|\tilde{n}_{k}\right|^2$ which is also $\beta$.

\subsection{The RG flow in the \boldmath{$zn$} model}

Now we can  study  the RG flow equation for the K\"ahler potential of the  $zn$ model. We will limit ourselves  to the $N=2$ case and show that the metric of the $zn$ model flows to the Ricci-flat conifold metric \eqref{24} in the IR. 

 The classical  K\"ahler potential of the $zn$ model \eqref{72} falls into the same class as the one of  $\mathbb{WCP}$, namely, it is described by the same {\em Ansatz} \eqref{17}, 
\beq
	K_{zn} = f(\tilde{r}^{2}) + \beta\log\left(1+\left|\frac{\tilde{n}_{1}}{\tilde{n}_{2}}\right|^{2}\right),
\eeq
where $f(\tilde{r}^{2})$ is  a function depending only on the radial coordinate $\tilde{r}^{2}\equiv \left|\zeta\right|^{2}$, while the second term is  the standard $\mathbb{CP}(N-1)$ K\"ahler potential. This suggests that we can use the {\em Ansatz} 
above to study the RG flow in the $zn$ model. The RG equation is essentially the same as in (\ref{218}), 
\beq
	\frac{\pd f}{\pd t} = \frac{1}{2\pi} \Big[ \log{f'} + \log(\beta+\tilde{r}^{2}f') + \log(f'+\tilde{r}^{2}f'')\Big] +C,
	\label{znRGeqnf}
\eeq
where now the prime denotes  derivatives with respect to $\tilde{r}^2$, while $C$
is an integration constant. The initial UV condition for $f^{{\rm UV}} (t=0,\tilde{r}^{2})$ is given by \eqref{72}, namely
\beq
f^{{\rm UV}} (t=0,\tilde{r}^{2}) = \tilde{r}^{2}.
\label{zninitialf}
\eeq

Much in  the same way as for the $\mathbb{WCP}(N,N)$ model we rewrite Eq. \eqref{znRGeqnf} in terms of the function $\gamma$,
\beq
\gamma \equiv \tilde{r}^{2} f'(\tilde{r}^{2}).
\label{gammazn}
\eeq
This gives the same equation as in \eqref{RGeqn}, namely
\beq
\frac{\pt \gamma}{\pt t} = \frac1{2\pi}\left\{r^2\,\frac{\gamma'^2(2\gamma +\beta) + \gamma(\gamma+\beta)\gamma''}{\gamma
(\gamma+\beta)\gamma'}-1\right\}.
\label{RGeqnzn}
\eeq
We  solve this equation numerically below with the initial condition
\beq
\gamma^{{\rm UV}} (t=0,\tilde{r}^{2}) = \tilde{r}^{2},
\label{zninitial}
\eeq
 see \eqref{zninitialf} and  the boundary condition of the form
\beq
\gamma'(t,\tilde{r}^{2}=L) = \gamma_{IR}'(t,\tilde{r}^{2}=L) .
\label{znbc}
\eeq

\begin{figure}[H]
	
	\begin{subfigure}[b]{0.5\linewidth}
		\includegraphics[width=\linewidth]{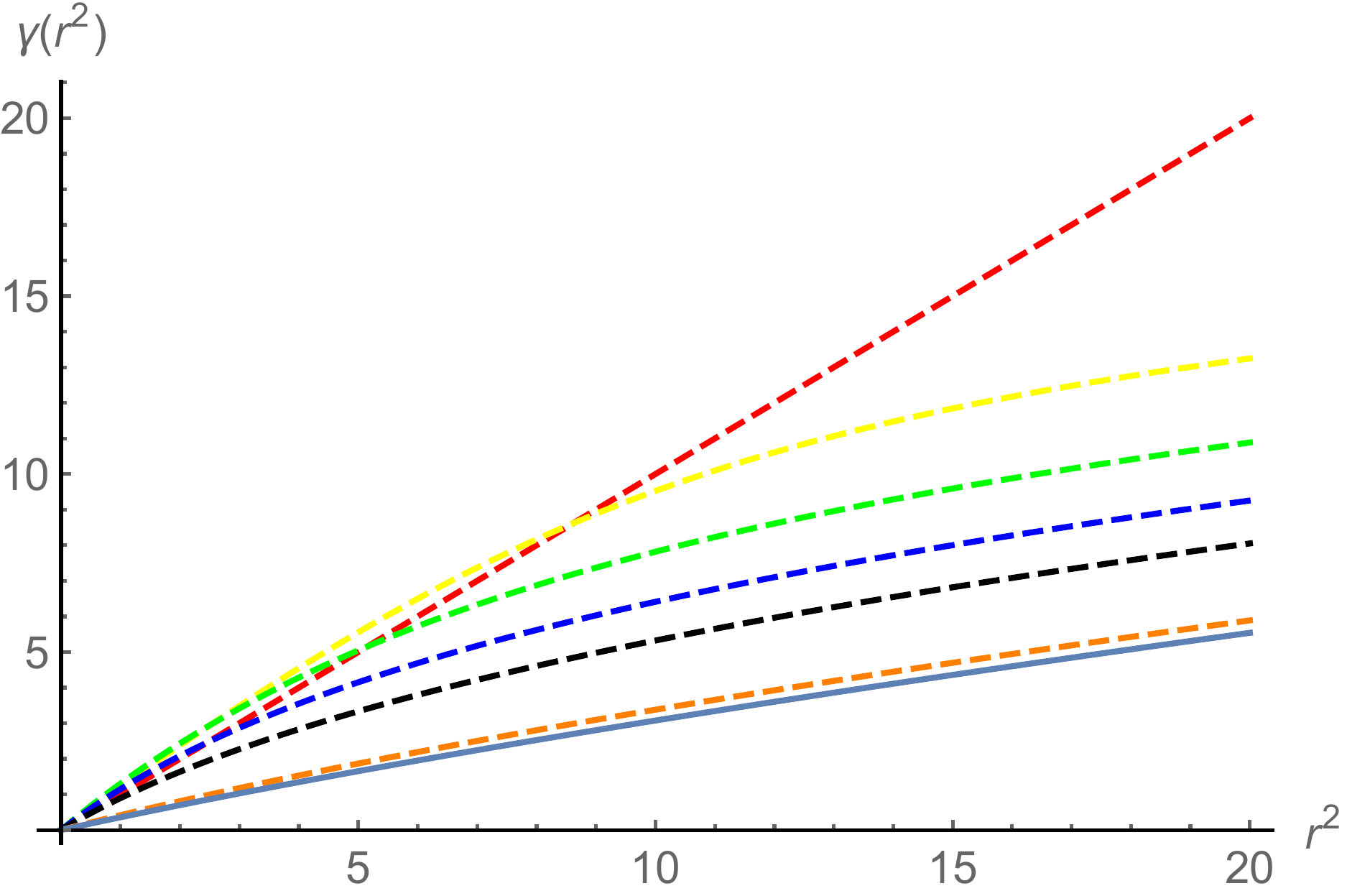}
		\caption{
$\beta=5$}
	\end{subfigure}
	\begin{subfigure}[b]{0.5\linewidth}
		\includegraphics[width=\linewidth]{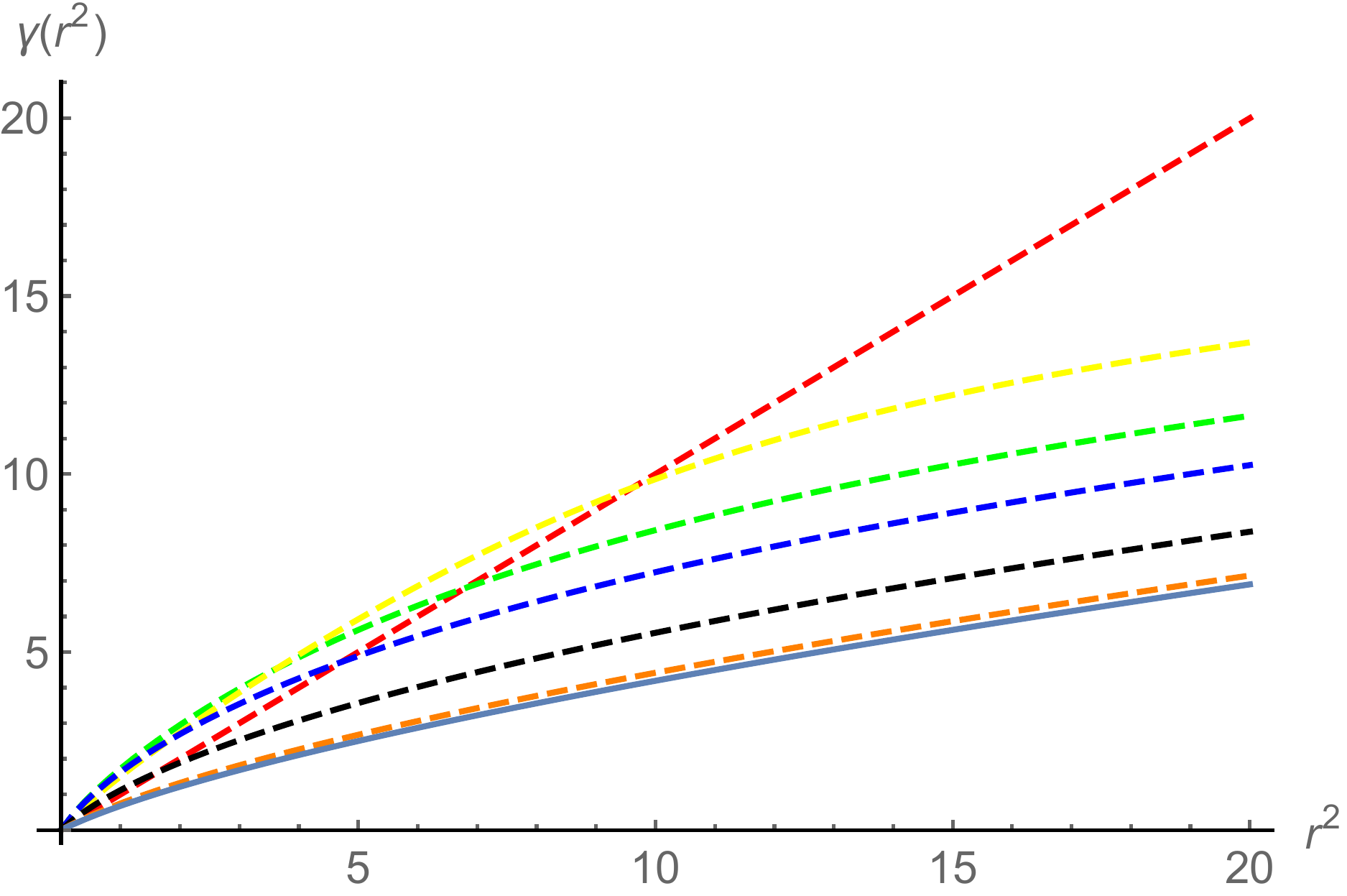}
		\caption{
$\beta=1$}
	\end{subfigure}
\center
	\begin{subfigure}[b]{0.5\linewidth}
\center
		\includegraphics[width=\linewidth]{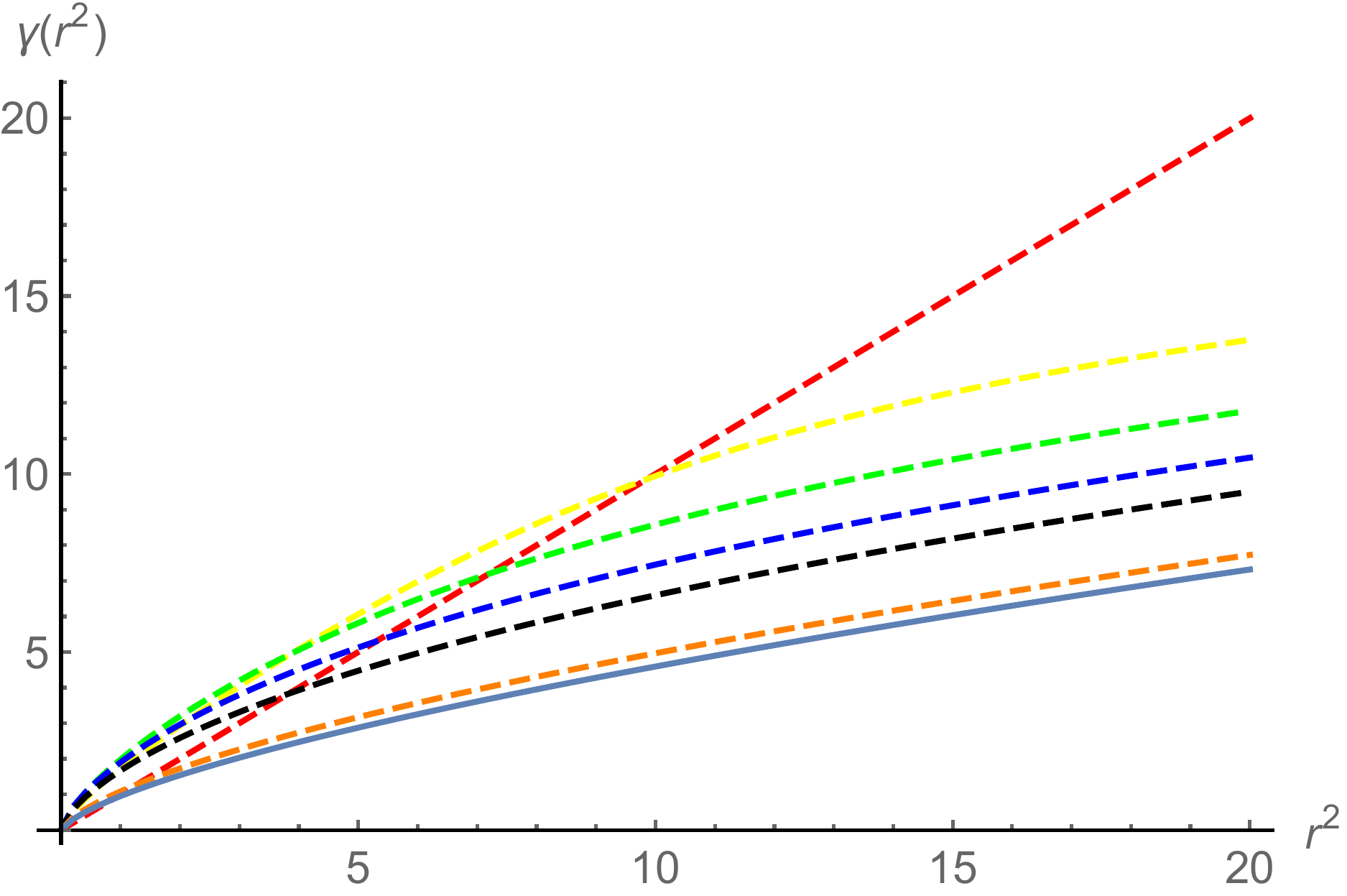}
		\caption{
$\beta=0.1$}
	\end{subfigure}

	\caption{\small Convergence of the IR flow of $\gamma(\tilde r^2)$ from the starting UV point $\gamma = \tilde r^2$ (red dashed line) to the analytic IR solution (solid blue line).  From top to bottom dashed lines (yellow to orange) represent the intermediate K\"ahler potentials from the early RG time to the late RG time. }
	\label{fignew1}
\end{figure}

\begin{figure}[H]
	
	\begin{subfigure}[b]{0.5\linewidth}
		\includegraphics[width=\linewidth]{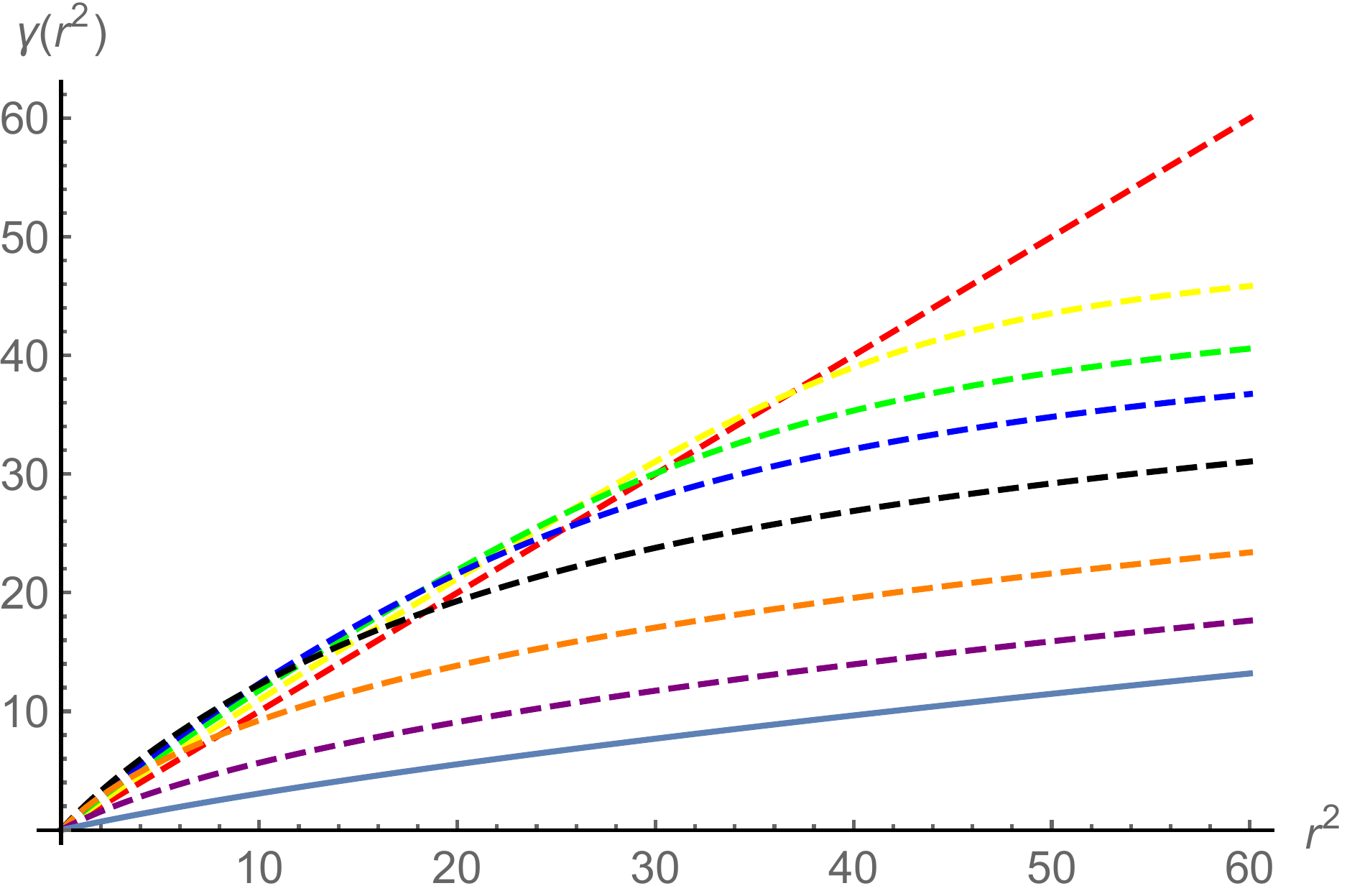}
		\caption{
$\beta=5$}
	\end{subfigure}
	\begin{subfigure}[b]{0.5\linewidth}
		\includegraphics[width=\linewidth]{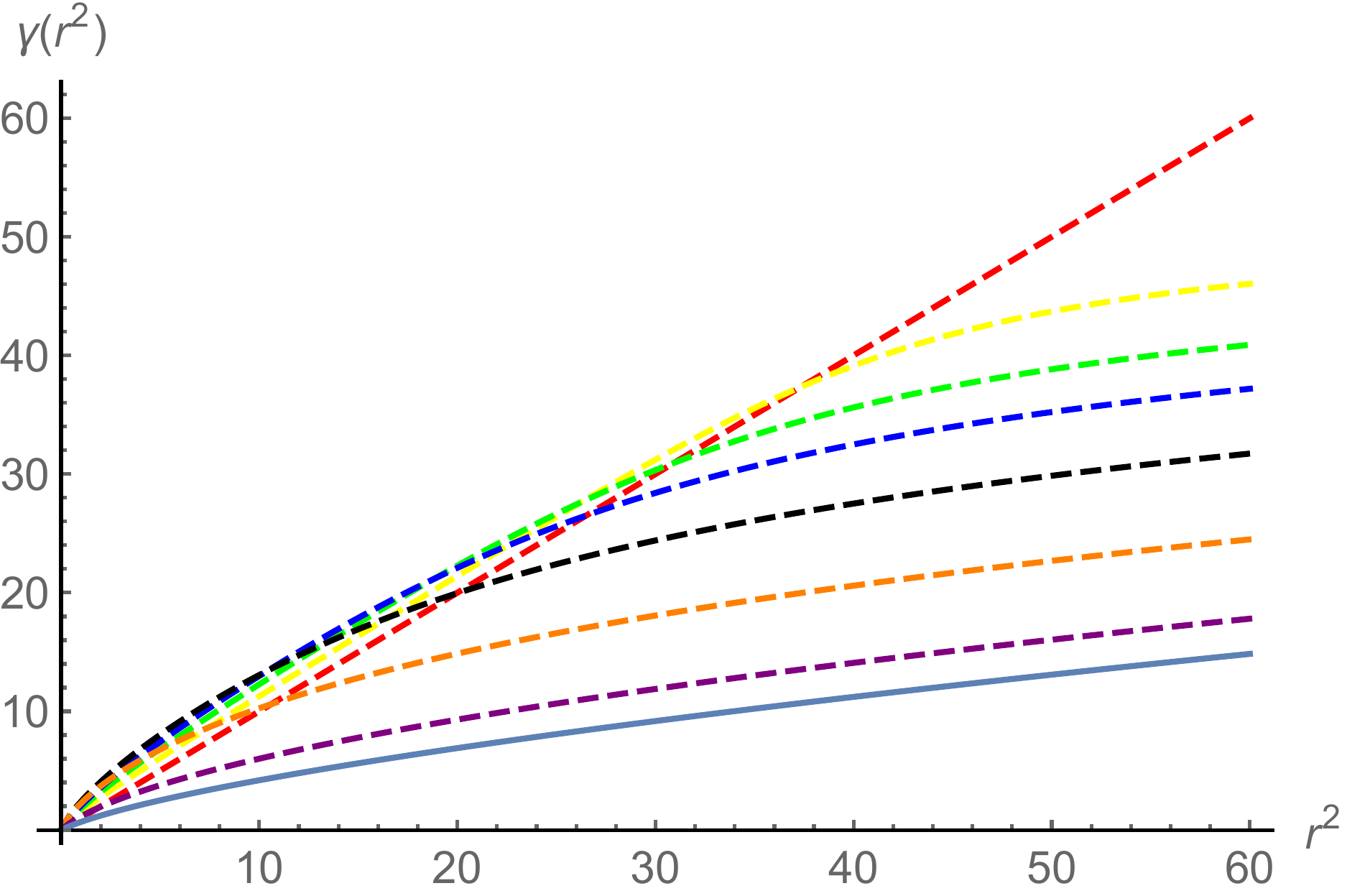}
		\caption{
$\beta=1$}
	\end{subfigure}
\center
	\begin{subfigure}[b]{0.5\linewidth}
\center
		\includegraphics[width=\linewidth]{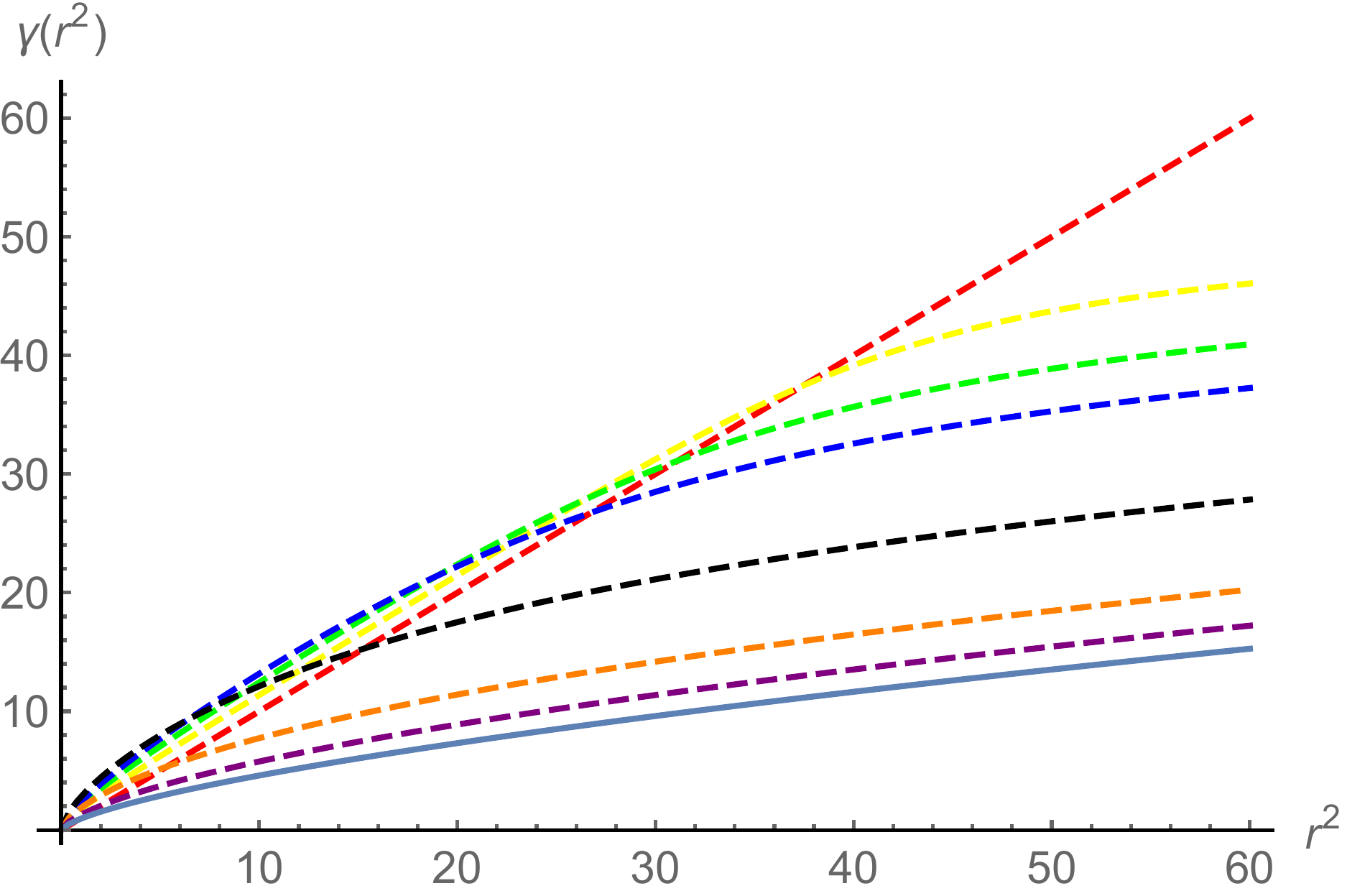}
		\caption{
$\beta=0.1$}
	\end{subfigure}

	\caption{\small Test at larger $\tilde r^2$ of convergence of the IR flow of $\gamma(\tilde r^2)$ from the starting UV point $\gamma = \tilde r^2$ (red dashed line) to the analytic IR solution (solid blue line).  From top to bottom dashed lines (yellow to purple) represent the intermediate K\"ahler potentials from the early RG time to the late RG time.}
	\label{fignew2}
\end{figure}

Figures \ref{fignew1} and \ref{fignew2} show the results of the UV to IR convergence for different values of $\beta$ and at different $L$. All results indicate a stable convergence of the UV starting point towards the IR solution. 
Note also that the exact twisted superpotential of the $zn$ model  coincides with the one in $\mathbb{WCP}(N,N)$ \cite{KS}. Thus, in much the same way as in $\mathbb{WCP}(N,N)$ we conclude that no dynamical mass gap is generated due to $\sigma$ VEV (which does not develop).
 
\section{Conclusions}

In this paper we thoroughly discussed the relationship between the GLSM and NLSM formulations of one and the same model referred to as in $\mathbb{WCP}(N,N)$.
The focus of our study was $\mathbb{WCP}(2,2)$.  Its GLSM formulation is equivalent to 2D SQED with four flavors: 2 of charge 1 and two of charge -1, and the FI term $\beta$. Both formulations lead to identical predictions in the IR, namely the six-dimensional (three complex dimensions) Calabi-Yau manifold as the target space of a superconformal sigma model. This is the so-called resolved conifold. The authors of \cite{Candelas:1989js} came to this conclusion 
from the analysis of GLSM, in an indirect but simple way. First, they observed ${\mathcal N}=2$ that requires a K\"ahler manifold. 
Second, they noted that with the given matter sector
$\beta$ is not renormalized, which implies Ricci-flatness and conformality. Third, they used the fact that global symmetries 
cannot be spontaneously broken in two dimensions. They also assumed constraint \eqref{coni}. Combining the above fact they came to conifold conclusion. An explicit expression for the metric was
obtained in \cite{PandoZayas:2000ctr}.

On the other hand, there is a standard procedure leading to NLSM. In the framework of this procedure one relies on the Higgs regime, assuming that some matter fields acquire
 vacuum expectation values which force Higgsing of the U(1) gauge boson. At large $\beta$ the vector superfield $V$ then becomes heavy and can be integrated out.
 After eliminating $V$ we arrive at a non-linear sigma model which does have logarithmic renormalizations ($\log M_V/\mu$ corrections) and is neither Ricci-flat nor conformal.
 The target space metric is rather contrived. Fortunately, the above renormalizations can be calculated order by order although the required procedure is  time and labor intensive.
 In this way one obtains RG equations which cannot be solved analytically, but only numerically.
 We analyzed the RG flow and demonstrated that the solution of the RG equations in the IR tends to the analytic metric of   \cite{PandoZayas:2000ctr}.
 The road to Ricci flatness is neither straightforward nor easy. 
 
 What is the reason? 
 
 The starting point in the NLSM formulation is far from the exact solution. It assumes Higgsing which in fact does not occur in the case at hand --
 in the final solution quantum-mechanical fluctuations smear the fields $n$ and $\rho$ all over the target space.
 The same is true not only for $\mathbb{WCP}(N,N)$ but even in more conventional $\mathbb{CP}(N-1)$ models. Passing to NLSM implies a non-linear realization of
 the global SU$(N)$ symmetry, while the exact solution (known at $N=2$ \cite{zam} and $N\gg1$ \cite{W79}) proves its linear realizations.

\section*{Acknowledgments}

The authors are grateful to Sergey Ketov  and David Tong  for useful  discussions.
This work  is supported in part by DOE grant DE-SC0011842. The work of J.C. was supported by the National
Thousand-Young-Talents Program of China. G.T. is funded by a Fondecyt grant no. 1200025.
The work of A.Y. was  supported by William I. Fine Theoretical Physics Institute,   
University of Minnesota and 
by Russian Foundation for Basic Research Grant No. 18-02-00048a.


\addcontentsline{toc}{section}{References}
\begin{thebibliography}{99}

\bibitem{HT1}
A.~Hanany and D.~Tong,
{\em Vortices, instantons and branes,}
JHEP {\bf 0307}, 037 (2003)
[hep-th/0306150].

\bibitem{ABEKY}
R.~Auzzi, S.~Bolognesi, J.~Evslin, K.~Konishi and A.~Yung,
 {\em Non-Abelian superconductors: Vortices and
 confinement in ${\mathcal N}=2$ 
 SQCD,}
Nucl.\ Phys.\ B {\bf 673}, 187 (2003)
[hep-th/0307287].

\bibitem{SYmon}
M.~Shifman and A.~Yung,
{\em Non-Abelian String Junctions as Confined Monopoles,}
Phys. Rev. D \textbf{70}, 045004 (2004)
[arXiv:hep-th/0403149 [hep-th]].

\bibitem{HT2}
A. Hanany and D. Tong,
{\em Vortex strings and four-dimensional gauge dynamics,}
JHEP {\bf 0404}, 066 (2004).
[hep-th/0403158].

\bibitem{Trev}
D.~Tong,
{\em TASI lectures on solitons: Instantons, monopoles, vortices and kinks,}
  hep-th/0509216.

\bibitem{Jrev}
  M.~Eto, Y.~Isozumi, M.~Nitta, K.~Ohashi and N.~Sakai,
{\em Solitons in the Higgs phase: The moduli matrix approach,}
  J.\ Phys.\ A  {\bf 39}, R315 (2006)
  [arXiv:hep-th/0602170].
  
  \bibitem{SYrev}
M.~Shifman and A.~Yung,
{\em Supersymmetric Solitons and How They Help Us Understand Non-Abelian Gauge Theories,}
  Rev.\ Mod.\ Phys.\  {\bf 79}, 1139 (2007)
  [hep-th/0703267]; for an expanded version see
{\sl Supersymmetric Solitons,}
(Cambridge University Press, 2009).

\bibitem{Trev2}
D.~Tong,
{\em Quantum Vortex Strings: A Review,}
  Annals Phys.\  {\bf 324}, 30 (2009)
  [arXiv:0809.5060 [hep-th]].
	
	\bibitem{SYsem}
 M.~Shifman and A.~Yung,
  {\em Non-Abelian semilocal strings in ${\mathcal N} = 2$ supersymmetric QCD,}
  Phys. Rev. D \textbf{73}, 125012 (2006)
[arXiv:hep-th/0603134 [hep-th]].
  
\bibitem{Jsem}
M.~Eto, J.~Evslin, K.~Konishi, G.~Marmorini, et al.,
 {\em On the moduli space of semilocal strings and lumps,}
  Phys.\ Rev.\  D {\bf 76}, 105002 (2007).
  
\bibitem{SVY}
 M.~Shifman, W.~Vinci and A.~Yung,
 {\em Effective World-Sheet Theory for Non-Abelian Semilocal Strings in N = 2 Supersymmetric QCD,}
  Phys. Rev. D \textbf{83}, 125017 (2011)
[arXiv:1104.2077 [hep-th]].
	
\bibitem{W79} 
E.~Witten,
{\em Instantons, the Quark Model, and the 1/N Expansion,}
  Nucl.\ Phys.\ B {\bf 149}, 285 (1979).

\bibitem{W2} 
 E.~Witten,
{\em Phases of N=2 Theories in Two Dimensions,}
  Nucl.\ Phys.\ B {\bf 403}, 159 (1993)
  [hep-th/9301042].

	\bibitem{HaHo}
A.~Hanany and K.~Hori,
  ``Branes and N = 2 theories in two dimensions,''
  Nucl.\ Phys.\  B {\bf 513}, 119 (1998)
  [arXiv:hep-th/9707192].

	\bibitem{V}
K.~Hori and C.~Vafa,
{\em Mirror Symmetry,}
  hep-th/0002222.
	
\bibitem{SYcstring} 
  M.~Shifman and A.~Yung,
 {\em Critical String from Non-Abelian Vortex in Four Dimensions,}
  Phys.\ Lett.\ B {\bf 750}, 416 (2015)
  [arXiv:1502.00683 [hep-th]].

 \bibitem{KSYconifold}
P.~Koroteev, M.~Shifman and A.~Yung,
{\em  Non-Abelian Vortex in
 Four Dimensions as a Critical  String on a Conifold},
 Phys.\ Rev.\ D {\bf 94} (2016) no.6,  065002
  [arXiv:1605.08433 [hep-th]].

\bibitem{SYlittles} 
  M.~Shifman and A.~Yung,
{\em Critical Non-Abelian Vortex in Four Dimensions and Little String Theory,}
  Phys.\ Rev.\ D {\bf 96}, no. 4, 046009 (2017)
  [arXiv:1704.00825 [hep-th]].

  \bibitem{NV}
  A.~Neitzke and C.~Vafa,
{\em Topological strings and their physical applications,}
  hep-th/0410178.
  
  \bibitem{KS}
  P.~Koroteev, M.~Shifman, W.~Vinci and A.~Yung,
{\em Quantum Dynamics of Low-Energy Theory on Semilocal Non-Abelian Strings,}
  Phys.\ Rev.\ D {\bf 84}, 065018 (2011)
  [arXiv:1107.3779 [hep-th]].
  
      \bibitem{SS}
C.~H.~Sheu and M.~Shifman,
{\em From Gauged Linear Sigma Models to Geometric Representation of $\mathbb{WCP}(N,\tilde{N})$ in 2D,}
  Phys.\ Rev.\ D {\bf 101}, no. 2, 025007 (2020)
  [arXiv:1907.09460 [hep-th]].
  
  \bibitem{ARSW}
O.~Aharony, S.~S.~Razamat, N.~Seiberg and B.~Willett,
  JHEP {\bf 1702}, 056 (2017)
  [arXiv:1611.02763 [hep-th]].
  
  \bibitem{Candelas:1989js}
  P.~Candelas and X.~C.~de la Ossa,
  {\em Comments on Conifolds,}
  Nucl.\ Phys.\ B {\bf 342}, 246 (1990).
  
	\bibitem{PandoZayas:2000ctr}
	L.~A.~Pando Zayas and A.~A.~Tseytlin,
	{\em 3-branes on resolved conifold,}
	JHEP \textbf{11}, 028 (2000)
	[arXiv:hep-th/0010088 [hep-th]].
	
	\bibitem{zam}
A.~B.~Zamolodchikov and A.~B.~Zamolodchikov,
{\em Factorized $S$ Matrices in Two-Dimensions as the Exact Solutions of Certain Relativistic Quantum Field Models,}
Annals Phys. \textbf{120}, 253-291 (1979).
	
\bibitem{Li}
	Chi~Li, 
	{\em On Rotationally Symmetric K\"ahler-Ricci Solitons}, 
	arXiv:1004.4049 [hep-th].
 
  
  
	\bibitem{AdDVecSal}
A.~D'Adda, A.~C.~Davis, P.~DiVeccia and P.~Salamonson,
Nucl.\ Phys.\ {\bf B222} 45 (1983).
	

  \bibitem{DHT}
  N.~Dorey, T.~Hollowood and D.~Tong,
  {\em The BPS Spectra of Gauge Theories in Two and Four Dimensions,}
  JHEP {\bf 9905}, 006, (1999)
  [hep-th/9902134].
   
  
  
\end{thebibliography}
\end{document}